\documentstyle[myaasmacros,alleqno,astrobib2,psfig,epsfig,times]{mn}

\newif\ifAMStwofonts

\ifoldfss
  \ifCUPmtlplainloaded \else
    \NewTextAlphabet{textbfit} {cmbxti10} {}
    \NewTextAlphabet{textbfss} {cmssbx10} {}
    \NewMathAlphabet{mathbfit} {cmbxti10} {} 
    \NewMathAlphabet{mathbfss} {cmssbx10} {} 
  \fi
  \ifAMStwofonts
    \ifCUPmtlplainloaded \else
      \NewSymbolFont{upmath} {eurm10}
      \NewSymbolFont{AMSa} {msam10}
      \NewMathSymbol{\upi}     {0}{upmath}{19}
      \NewMathSymbol{\umu}     {0}{upmath}{16}
      \NewMathSymbol{\upartial}{0}{upmath}{40}
      \NewMathSymbol{\leqslant}{3}{AMSa}{36}
      \NewMathSymbol{\geqslant}{3}{AMSa}{3E}

      \let\leq=\leqslant \let\le=\leqslant
       
    \fi
  \fi
\fi 

\ifnfssone
  \newmathalphabet{\mathit}
  \addtoversion{normal}{\mathit}{cmr}{m}{it}
  \addtoversion{bold}{\mathit}{cmr}{bx}{it}
  \newmathalphabet{\mathbfit} 
  \addtoversion{normal}{\mathbfit}{cmr}{bx}{it}
  \addtoversion{bold}{\mathbfit}{cmr}{bx}{it}
  \newmathalphabet{\mathbfss} 
  \addtoversion{normal}{\mathbfss}{cmss}{bx}{n}
  \addtoversion{bold}{\mathbfss}{cmss}{bx}{n}
  \ifAMStwofonts
    \ifCUPmtlplainloaded \else
      \UseAMStwoboldmath
      \makeatletter
      \new@mathgroup\upmath@group
      \define@mathgroup\mv@normal\upmath@group{eur}{m}{n}
      \define@mathgroup\mv@bold\upmath@group{eur}{b}{n}
      \edef\UPM{\hexnumber\upmath@group}
      \new@mathgroup\amsa@group
      \define@mathgroup\mv@normal\amsa@group{msa}{m}{n}
      \define@mathgroup\mv@bold\amsa@group{msa}{m}{n}
      \edef\AMSa{\hexnumber\amsa@group}
      \makeatother
      \mathchardef\upi="0\UPM19
      \mathchardef\umu="0\UPM16
      \mathchardef\upartial="0\UPM40
      \mathchardef\leqslant="3\AMSa36
      \mathchardef\geqslant="3\AMSa3E

      \let\leq=\leqslant \let\le=\leqslant

    \fi
  \fi
\fi 

\ifnfsstwo
  \DeclareMathAlphabet{\mathbfit}{OT1}{cmr}{bx}{it}
  \SetMathAlphabet\mathbfit{bold}{OT1}{cmr}{bx}{it}
  \DeclareMathAlphabet{\mathbfss}{OT1}{cmss}{bx}{n}
  \SetMathAlphabet\mathbfss{bold}{OT1}{cmss}{bx}{n}
  \ifAMStwofonts
    \ifCUPmtlplainloaded \else
      \DeclareSymbolFont{UPM}{U}{eur}{m}{n}
      \SetSymbolFont{UPM}{bold}{U}{eur}{b}{n}
      \DeclareSymbolFont{AMSa}{U}{msa}{m}{n}
      \DeclareMathSymbol{\upi}{0}{UPM}{"19}
      \DeclareMathSymbol{\umu}{0}{UPM}{"16}
      \DeclareMathSymbol{\upartial}{0}{UPM}{"40}
      \DeclareMathSymbol{\leqslant}{3}{AMSa}{"36}
      \DeclareMathSymbol{\geqslant}{3}{AMSa}{"3E}

      \let\leq=\leqslant \let\le=\leqslant

    \fi
  \fi
\fi 

\ifCUPmtlplainloaded \else
  \ifAMStwofonts \else 
    \def\upi{\pi}
    \def\umu{\mu}
    \def\upartial{\partial}
  \fi
\fi

\title[Mass loss from dSph galaxies]{Mass loss from dwarf spheroidal
  galaxies: the origins of 
  shallow dark matter cores and exponential surface brightness
  profiles}
\author[J. I. Read and G. Gilmore]
       {J. I. Read and G. Gilmore \\
        Institute of Astronomy, Cambridge University, Madingley Road, Cambridge, CB3 0HA}
\date{Accepted.
      Received;
      in original form}

\pagerange{\pageref{firstpage}--\pageref{lastpage}}
\pubyear{2003}

\begin{document}

\maketitle

\label{firstpage}

\begin{abstract}
Dwarf spheroidal galaxies have shallow central dark matter density
profiles, low angular momentum and approximately exponential surface
brightness distributions. Through N-body simulations and analytic
calculations we investigate the extent to which these properties can
be generated from ``typical'' $\Lambda$CDM galaxies, which differ in
all of these properties, by the dynamical consequences of feedback.

We find that, for a wide range of initial conditions, one impulsive
mass loss event will naturally produce a surface brightness profile in
the remaining stellar component of a dwarf
spheroidal galaxy (dSph) which is well fit over many scale lengths by an
exponential, in good qualitative agreement with observations of
Local Group dSphs. Furthermore, two impulsive mass loss phases,
punctuated by significant gas re-accretion, are
found to be sufficient to transform a central density cusp in the dark
matter profile into a near-constant density core. This may then
provide the missing link between current cosmological simulations,
which predict a central cusp in the dark matter density profile, and
current observations, which find much shallower central density
profiles.

We also look at the angular momentum history of dSphs and demonstrate
that if these galaxies have spent most of their lifetime in tidal isolation
from massive galaxies then they cannot have formed from high angular
momentum gas discs.

\end{abstract}


\section{Introduction}\label{sec:introduction}
Over recent years, dwarf galaxies have become increasingly interesting
objects in the 
light of hierarchical formation models. In the current $\Lambda$CDM
paradigm, all structure in the universe forms from the successive
mergers of smaller substructures (see e.g. \citet{1978MNRAS.183..341W}, 
\citet{1996ApJ...462..563N}, \citet{2000ApJ...538..528S},
\citet{2002Astro-Ph..0205448}, \citet{2004ApJ...607..688G}; or for a review
\citet{1999Ap&SS.269..513S}). In this scenario, low mass dwarf
galaxies may be the left-over remnants of the hierarchical merging
process. If this is the case, then their stellar populations
should be able to tell us much about merger histories
\citep{2003AJ....125..707T}; their internal angular momentum and surface
brightness profiles should contain vital clues as to the coupling of
the gas to the dark matter in the early universe
\citep{2001MNRAS.326.1205V}; and their underlying
dark matter distribution, largely unaffected by merging, should be able
to tell us much about the detailed nature of dark matter
\citep{2001ApJ...563L.115K}.

The dwarf galaxies of the Local Group can be split into three broad 
categories: dwarf irregular galaxies (dIrr), which are rich in HI gas,
transition galaxies (dIrr/dSph), which contain some HI gas and dwarf
spheroidal galaxies (dSph), which contain little or no HI gas
(\citet{1998ARA&A..36..435M}, \citet{1999A&ARv...9..273V},
\citet{1994A&ARv...6...67F} and \citet{1999IAUS..192...17G}). The dSph
galaxies have surface brightness profiles which are well fit by an
exponential over many scale lengths (\citet{2000Astro-Ph..0207467} and
\citet{2001AJ....122.2538O}) and there is
mounting evidence that many of the Local Group dwarfs have very little
internal angular momentum (\citet{2001ApJ...563L.115K} and
\citet{1998ARA&A..36..435M}). Finally, while little is known in general
about the distribution of dark matter in these galaxies, one recent
indirect measurement suggests that Ursa Minor (UMi) has a central,
constant density, dark matter core \citep{2003ApJ...588L..21K},
similar to those observed in LSB galaxies (see
e.g. \citet{2001ApJ...552L..23D}).

In contrast to these observations, if the dwarf galaxies of the Local
Group are indeed the untouched 
remnants of a hierarchical merging process, then current numerical
simulations predict a wide range of angular momenta for these
galaxies (see e.g. \citet{2001ApJ...555..240B} and
\citet{Colin:2003jd}), while their central dark matter density
distribution should exhibit a log-slope of $1 < \alpha < 1.5$ (see e.g
\citet{2001ApJ...554..903K}, \citet{2000ApJ...544..616G} and
\citet{1996ApJ...462..563N}).

These differences between the numerical experiments and the
observations could be an indication that the current cosmological
paradigm breaks down on small scales; or it could be a result of a
complex evolutionary history for the Local Group dwarfs due to some form
of {\it feedback}. The
evidence for a complex evolution is compelling: gas poor dSph galaxies
are found preferentially close to their host galaxy while gas rich
dIrr lie further away, indicating an environmental dependence for the
galaxy morphology (\citet{1998ARA&A..36..435M},
\citet{2003AJ....125.1926G} and \citet{1999A&ARv...9..273V}); the star
formation histories for the Local Group dSph galaxies are complex and
varied (\citet{2002MNRAS.332...91D}, \citet{2003AJ....126..218M},
\citet{2000MNRAS.317..831H} and \citet{2002A&A...391...55I}); and the
currently very low mass to light ratio, small HI mass, and low
chemical abundance in dSph
galaxies is suggestive of significant gas mass loss
(\citet{1998ARA&A..36..435M}, \citet{2003AJ....125.1926G},
\citet{2001ApJ...563L.115K} and \citet{2002MNRAS.334..117C}).

From a theoretical viewpoint, it has been known for some time that feedback
is required within hierarchical clustering theories in order to
explain the lack of small scale power in the observed galaxy
luminosity function (\citet{1978MNRAS.183..341W},
\citet{2003ApJ...599...38B} and \citet{2004MNRAS.347.1093B}); while
mass loss as a form of feedback has been discussed extensively in the
literature, both with regards to large spiral galaxies (see
e.g. \citet{1998A&A...331L...1S}) and
smaller dwarf galaxies (see e.g. \citet{1974MNRAS.169..229L},
\citet{1986ApJ...303...39D}, 
\citet{2003Astro-Ph..0210454}, \citet{2003ApJ...584..541H} and
\citet{2000MNRAS.317..697E}).

Many mechanisms have been proposed to drive gas mass loss from dwarf
galaxies, including ram pressure stripping (\citet{1974Natur.252..111E} and
\citet{2003AJ....125.1926G}), supernovae explosions
(\citet{1974MNRAS.169..229L}, \citet{1986ApJ...303...39D},
\citet{2003Astro-Ph..0210454} and \citet{2000MNRAS.317..697E});
photoevaporation (\citet{1996MNRAS.278L..49Q} and
\citet{1999ApJ...523...54B}); and the effects
of galaxy harassment and tidal stripping from a large host galaxy
(\citet{2001ApJ...547L.123M}, \citet{1996Natur.379..613M}, 
\citet{1998ApJ...495..139M} and \citet{2003ApJ...584..541H}). It is
likely that all of these mechanisms act in 
tandem to a greater or lesser extent in the evolution of both dSph and
dIrr galaxies. 

In this study we consider the dynamical effect of mass loss on the
remaining stars and dark matter in isolated dSph galaxies. We simplify
the analysis to be purely dynamical, explicitly not considering the
physical processes which drive the mass loss. In this way we are able
to study the maximum possible dynamical effect on the surviving
galaxy, to see if, in principle, extreme mass loss can change the
structure of dwarf galaxies.

Other authors have previously studied the dynamical effect of mass
loss from dwarf galaxies but focused only on the effects on the
remaining dark matter (\citet{1996MNRAS.283L..72N},
\citet{1999MNRAS.303..321G} and \citet{2002MNRAS.333..299G}), or
specifically on the hydrodynamics of gas mass loss
(\citet{1999ApJ...513..142M}).

In this paper we present the first comprehensive dynamical study of
mass loss from progenitors to what are currently dSph galaxies,
using the fully self consistent 3D N-body 
code, GADGET \citep{2001NewA....6...79S}. We perform the highest
resolution simulations to date and unlike previous studies we
focus both on the response of the dark matter to mass loss {\it and}
the response of the stellar component. This last point is critical
since the final surface brightness and angular momentum distribution
of the stars in our dynamical models can then be compared with data
from the Local Group dSph galaxies.

In doing this we pose the question: can mass loss explain
the low angular momentum, exponential surface brightness distribution
and cored central dark matter distribution observed in the dSph
galaxies of the Local Group?

This paper is organised as follows: In section \ref{sec:nummethod}, we
present our numerical method for setting up the initial conditions,
performing the mass loss and subsequently evolving the system. In
section \ref{sec:results}, we 
present the results of a suite of numerical N-body simulations to
determine the dynamical response of a two-component galaxy comprising
baryons\footnote{We refer throughout this paper to the gas and
  stars collectively as baryons. This is because we do not include any
  gas hydrodynamics in our models, and so the gas and stars are
  dynamically indistinguishable.} and dark matter, initially in
equilibrium, to significant 
baryonic mass loss. This section is divided into three sub-sections,
each focusing on a potential observational signature of the remnant
galaxy after mass loss. Section \ref{sec:resangmom} focuses on the
dynamical effect of mass loss on the angular momentum distribution of
the baryons, section \ref{sec:resdarkmatter} focuses on the dynamical
effect of mass loss on the radial density profile of the dark matter
and section \ref{sec:ressurfden} focuses on the surface density
profile of the baryons after mass loss. In section
\ref{sec:discussion}, we discuss the simulation results in comparison
with other results already in the literature and highlight, clarify
and justify the choice of assumptions inherent in this work. In
section \ref{sec:observations}, we then go on to compare the results
of the simulations with observations of dwarf galaxies from the Local
Group. Finally, in section \ref{sec:conclusions}, we present our
conclusions.

\section{The numerical method}\label{sec:nummethod}

In this section we describe the methodology and initial conditions
for a suite of simulations which track the dynamical response of a
two component system of baryons and dark matter to significant baryon
mass loss.

\subsection{The initial conditions}\label{sec:initialcond}
We used a two-component model for the initial conditions comprising some
baryons and some dark matter. The system was set up in equilibrium as in
\citet*{1993ApJS...86..389H}. The density profiles were populated with
particles using Monte Carlo methods, and the velocity structure was
then realised using moments of the Collisionless Boltzmann Equation,
assuming a Maxwellian distribution (see \citet{1993ApJS...86..389H}
for more detail).

A recent paper by \citet{2004ApJ...601...37K} has suggested that the
Maxwellian velocity approximation used by \citet*{1993ApJS...86..389H} in
setting up the initial conditions can lead to spurious results for
simulations performed over long timescales. To test that the results
presented here are not affected by the Maxwellian approximation we
re-simulated one run using an initial condition generator which drew
the velocities directly from a numerically calculated distribution
function, as in \citet{2004ApJ...601...37K}. The results showed 
excellent agreement with the runs performed using the Maxwellian
approximation for the initial conditions.

We used two different density profiles for the baryons and two for the
dark matter in order to cover the extrema initial conditions of
interest. For the baryons, we considered a rotating exponential disc or
a generalised Hernquist spheroid, which may be set rotating if
required\footnote{As discussed in \citet{1993ApJS...86..389H}, the
  rotating spheroids start with systematically less angular momentum
  than the rotationally supported discs.}(see \citet{1990ApJ...356..359H},
\citet{1992MNRAS.254..132S} and \citet{1996MNRAS.278..488Z}). These
two profiles are relevant since an exponential disc may be
morphologically close to the dIrr of the Local Group at the present
epoch, while a spheroidal profile is close to the dSph galaxies of
the Local Group observed at the current epoch
(\citet{1998ARA&A..36..435M}, \citet{1999IAUS..192...17G} and
\citet{1999A&ARv...9..273V}). For the dark matter, we used one of two
profiles - the Hernquist profile, or the truncated isothermal sphere
\citep{1993ApJS...86..389H}. The Hernquist profile is
relevant since in the central regions it closely resembles the profiles
found in large N-body numerical simulations of dark matter structure
formation (see e.g. \citep{1996ApJ...462..563N} and
\citet{2000ApJ...544..616G}). The truncated isothermal sphere
profile is relevant since it provides a good fit to the dark matter
density profiles derived from real data (see
e.g. \citet{2001MNRAS.323..285B} and \citet{2001ApJ...552L..23D}). The
analytic forms for these profiles are summarised for completeness
below:

The exponential disc is given by:
\begin{equation}
\rho_d(R,z)=\frac{M_d}{4\pi h_d^2z_0}\exp(-R/h_d){\mathrm{sech}}^2(\frac{z}{z_0})
\label{eqn:expdisc} 
\end{equation}
where $R$ and $z$ are the familiar cylindrical coordinates, $h_d$ is the
disc scale length, $z_0$ is the disc scale height and $M_d$ is the
disc mass. 

The generalised Hernquist profile is given by (see
e.g. \citet{1990ApJ...356..359H}, \citet{1992MNRAS.254..132S} and
\citet{1996MNRAS.278..488Z}):

\begin{equation}
\rho_{b,i}(r) =
\frac{C(\alpha,\beta,\gamma)}{\left(\frac{r}{h_b}\right)^\alpha \left(1+\left(\frac{r}{h_b}\right)^\gamma\right)^{\frac{\beta-\alpha}{\gamma}}}
\label{eqn:spheroid}
\end{equation}

Where $C(\alpha,\beta,\gamma)$ is a normalisation constant, $h_b$ is
the baryon scale length, $\alpha$ is the power law log-slope of the
baryons interior to $h_b$, $\beta$ is the log-slope exterior to $h_b$
and $\gamma$ controls the smoothness of the transition at $h_b$.

The Hernquist profile is given by (c.f. equation \ref{eqn:spheroid}
with $\alpha=1$, $\beta=4$ and $\gamma=1$):

\begin{equation}
\rho_h(r) = \frac{M_h}{2\pi a_h^3}\frac{1}{r/a_h(1+r/a_h)^3} \\
\label{eqn:hernquist}
\end{equation}

where $M_h$ is the mass of the halo, and $a_h$ is the scale
length.

The truncated isothermal sphere is given by:

\begin{eqnarray}
\rho_h(r) & = &
\frac{M_h}{2\pi^{3/2}}\frac{\zeta}{r_t}\frac{\exp(r^2/r_t^2)}{r^2+r_c^2}
 \nonumber \\
\zeta & = & \left(1-\sqrt{\pi}q\exp(q^2)(1-\mathrm{erf}(q))\right)^{-1}
\label{eqn:isothermal}
\end{eqnarray}

where $M_h$ is the halo mass, $r_t$ is the tidal cut-off radius,
$r_c$ is the core radius and $q=r_c/r_t$.

For some of the runs we model a dissipation and collapse
phase for the baryons. For these runs the
baryon collapse was modelled either numerically, by slowly increasing
the mass of the baryons over the contraction time, $t_{cont}$, while
holding the baryons fixed (c.f. \citet{1999MNRAS.303..321G} and
\citet{2002ApJ...571L..89J}), or analytically using a similar
prescription to that set out in 
\citet{2002MNRAS.333..299G} (see appendix \ref{sec:zhaocalc}). For the
numerical baryon contraction, the system was then evolved for a
further 30 time units before mass loss to ensure equilibrium; the
results were found not to be sensitive to this parameter.

\subsection{Units}\label{sec:units}
For all of the simulations discussed subsequently, we use a system of
units such that $G=M_{d,b}=h_{d,b}=1$, where $M_{d,b}$ is the mass of
the baryons (disc or spheroid), $h_{d,b}$ is the scale length of the
baryons and $G$ is the gravitational constant. For initial conditions
which approximate a typical dIrr galaxy, $M_{d,b} \sim 10^8$M$_\odot$
and $h_{d,b} \sim 1$kpc. Unit time is then $4.7 \times 10^7$years, while
unit velocity is $20.7$kms$^{-1}$. From here on, we will leave the
results in simulation units so that they can be scaled for comparison
with a range of systems.

\subsection{Equilibrium tests and resolution issues}\label{sec:eqtests}
As with any simulation, it is important to asses which aspects of the
simulation can be viewed as physical and which are a product of finite
resolution and computational techniques. As such, we have been careful
to test the code using a wide range of softening parameters and
particle numbers. Equilibrium tests were carried out on every run for
15 time units to ensure that the system remained unchanged before
any mass was removed. Most runs were carried out with 15,000 disc and
150,000 halo particles and with disc and halo softening parameters of
0.04 and 0.02 respectively. The softening parameters were chosen using
the analytic criteria of \citet{2003MNRAS.338...14P}. Test runs with
lower softening parameters were compared with the standard runs and no
changes were observed. Finally, key runs were re-simulated at higher
resolution with 68,181 disc and 681,818 halo particles with softening
parameters of 0.02 and 0.011 respectively, to test for convergence;
excellent agreement was found with the lower resolution runs.

Poisson noise was found to set in at less than 0.1 baryon scale
lengths for all of the simulations, indicating that all simulations were
well resolved over the scales of interest for this study.

The N-body numerical calculations were performed using the GADGET tree
code (\citet{2001NewA....6...79S}). The GADGET code is particularly
advantageous for this study because:\\
\\
1. It allows for variable time steps for each particle and hence
provides excellent time resolution at small radii where we are
most interested in the halo density profile.\\
\\
2. It comes in a massively parallel distribution which scales very
efficiently with particle number, allowing us to re-simulate the most
interesting runs at high resolution.\\
\\
The GADGET code was found to conserve energy to better than one part
in $10^5$ over the typical simulation time of 15 time units.

As a further test of the GADGET code, we re-simulated one run using a
different Poisson integrator described in
\citet{2000ApJ...536L..39D}. For this run, we used fixed time steps
and evolved the particles with a standard leap-frog algorithm (see
e.g. \citet{1986Natur.324..446B}). We found excellent agreement with
the results from GADGET.

Finally, for a collisionless system, it is important that two-body
relaxation is negligible over the entire simulation time
(\citet{1987gady.book.....B} and \citet{2003MNRAS.338...14P}). For all
of the runs, two body relaxation became important only interior to
$\sim 2$ halo softening lengths over the longest simulation times of
100 units ($\sim 7$Gyrs). 

\subsection{Mass loss and system evolution}\label{sec:massloss}
We removed the baryonic mass by adding a large
velocity vector of random direction to some of the baryons such that a
fraction of them were removed over a given time period and from a
particular region of the initial disc or spheroid. The response due to
this mass loss was then tracked using the GADGET tree code (see
section \ref{sec:eqtests} and \citet{2001NewA....6...79S}).

It is important to note explicitly that there is {\it no gas physics}
in our model; we wish to model the {\it dynamical effects} of
mass loss, not the hydrodynamics involved in actually performing gas
mass removal.

After mass loss most runs were evolved for a further 15 time units
($\sim 1$Gyr). Two key runs were evolved for a longer time of 100
units ($\sim 7$Gyrs) to test for equilibrium. Given that the typical
orbital times of Local Group dSph galaxies around the Milky Way are
$\sim 1-2$Gyrs (\citet{Dinescu:2004pe}, \citet{2002AJ....124.3198P},
\citet{1996AAS...188.0901S} and \citet{1995AJ....110.2747S}), we do
not generally consider the long-time evolution of the models presented
here since it is likely that, over these longer times, tidal effects become
important. The longer-time evolution of these dynamical models,
including the effects of tides, will be the subject of future work.

\section{Results}\label{sec:results}

In this section we describe the results from a suite of simulations
which track the dynamical response of a two-component system
comprising baryons and dark matter, initially in equilibrium, to
significant baryon mass loss. The results are divided into
three sub-sections, each focusing on a potential observational
signature of the remnant galaxy after mass loss. Section
\ref{sec:resangmom} focuses on the dynamical effect of mass loss on the
angular momentum distribution of the baryons, section
\ref{sec:resdarkmatter} focuses on the dynamical effect of mass loss on
the radial density profile of the dark matter and section
\ref{sec:ressurfden} focuses on the surface density profile of the
baryons after mass loss. In all cases comparisons with real data are
deferred until section \ref{sec:observations}.

The simulations are labelled in order of discussion from A through to
C. The particle number, force softening, mass loss parameters and
initial condition density and velocity profile parameters are given in
table \ref{tab:initcond}. Some animations, illustrating the
simulations in action can be viewed at:
\begin{verbatim}www.ast.cam.ac.uk/~jir22\end{verbatim}

\begin{table*}
\begin{center}
\setlength{\arrayrulewidth}{0.5mm}
\begin{tabular}{llllllllll}
\hline
{\it Run} & {\it $N_b$} & {\it $\epsilon_b$} & {\it $N_{dm}$} & {\it
  $\epsilon_{dm}$} & {\it $t_{cont}$} & {\it Baryon Profile} & {\it
  Halo Profile} & {\it $\delta$} & {\it $t_{ml}$} \\
\hline
A1 & 68,181 & 0.02 & 681,818 & 0.011 & 0 & D,1,1,0.2,R & H,10,10 & 0.95 &
1 \\
A2 & 15,000 & 0.04 & 150,000 & 0.02 & 0 & D,1,1,0.2,R & H,2,10 & 0.95 &
1 \\
A3 & 15,000 & 0.04 & 150,000 & 0.02 & 0 & D,1,1,0.2,R & T,10,40,10 & 0.95 &
1 \\
A4 & 68,181 & 0.02 & 681,818 & 0.011 & 0 & S,1,4,1,R & H,10,10 & 0.95 &
1 \\
A5 & 15,000 & 0.04 & 150,000 & 0.02 & 0 & S,1,4,1,R & H,2,10 & 0.95 &
1 \\
A6 & 15,000 & 0.04 & 150,000 & 0.02 & 0 & S,1,4,1,R & T,10,40,10 & 0.95 &
1 \\
A7 & 68,181 & 0.02 & 681,818 & 0.011 & 0 & D,1,1,0.2,R & H,10,10 & 0.99 &
1 \\
A8 & 68,181 & 0.02 & 681,818 & 0.011 & 0 & S,1,4,1,R & H,10,10 & 0.97 &
1 \\
\hline
B1 & 15,000 & 0.04 & 150,000 & 0.02 & \(\begin{array}{l}(a) 1\\(b)
  15\\(c) \infty \end{array}\) & S,1,4,1 & H,10,10 & 0.95 &
1 \\
B2 & 68,181 & 0.02 & 681,818 & 0.011 & \(\begin{array}{l}(a) 15\\(b)
  40\\(c) \infty \end{array}\) & S,1,4,1 & H,10,50 & 0.95 &
1 \\
B3 & 15,000 & 0.04 & 150,000 & 0.02 & \(\begin{array}{l}(a) 15\\(b)
  \infty \end{array}\) & S,1,4,1 & H,2,10 & 0.95 &
1 \\
B4 & 68,181 & 0.02 & 681,818 & 0.011 & * & S,1,4,1 & H,5,10 & 0.95 &
1 \\
\hline
C1 & 68,181 & 0.02 & 681,818 & 0.011 & 0 & S,1,4,1 & H,10,10 & 0.95 &
1 \\
C2 & 15,000 & 0.04 & 150,000 & 0.02 & 0 & S,0,3,1 & H,2,10 & 0.95 &
1 \\
C3 & 15,000 & 0.04 & 150,000 & 0.02 & 0 & S,1,4,1 & {\bf Fix} H,10,10 & 0.95 &
1 \\
C4 & 15,000 & 0.04 & 150,000 & 0.02 & $\infty$ & S,1,4,1 & H,10,10 & 0.95 &
1 \\
C5 & 15,000 & 0.04 & 150,000 & 0.02 & 0 & S,1,4,1 & H,10,10 & 0.95 &
9 \\
\hline
\end{tabular}
\end{center}
\caption[Initial conditions for simulation runs A-C]{Initial
  conditions for simulation runs A-C. The table columns from left to
  right are: the run label, the number of baryon particles, $N_b$, the
  baryon force softening, $\epsilon_b$, the number of dark matter
  particles, $N_{dm}$, the dark matter force softening,
  $\epsilon_{dm}$, the baryon infall time, $t_{cont}$, the baryon
  radial density profile, the halo radial density profile, the baryon
  mass loss fraction, $\delta$, and finally, the baryon outflow time,
  $t_{ml}$. Where the dark matter was not allowed to contract in
  response to the addition of the baryons, $t_{cont}=0$, and where the
  contraction was set up analytically for the initial conditions (the
  true adiabatic case), $t_{cont}=\infty$ (see appendix
  \ref{sec:zhaocalc} for more
  details). The notation used for the 
  baryon density profile is: D,$M_b$,$h_d$,$z_0$,R for an exponential
  disc of mass, $M_b$, scale length, $h_d$, and scale height, $z_0$,
  which is [R]otationally supported (see equation \ref{eqn:expdisc}) and:
  S,$\alpha$,$\beta$,$\gamma$,[R] for a generalised Hernquist spheroid
  with inner log-density slope, $\alpha$, outer slope $\beta$,
  transition smoothing, $\gamma$, mass $M_b=1$ and potentially set in
  [R]otation initially (see equation \ref{eqn:spheroid}). The notation
  used for the halo density profile is: H,$M_h$,$a_h$ for a Hernquist
  spheroid of mass, $M_h$, and scale length, $a_h$, (see equation
  \ref{eqn:hernquist}) and T,$M_h$,$r_t$,$r_c$ for a truncated
  isothermal sphere of mass $M_h$, tidal radius, $r_t$ and core size
  $r_c$ (see equation \ref{eqn:isothermal}). For run C3, the halo was
  held fixed throughout the mass loss and not allowed to respond. The
  runs A1-A8 are all concerned with the effects of mass loss on the
  angular momentum profile of the baryons. B1-B4 are concerned with
  the effects of mass 
  loss on the underlying dark matter distribution and C1-C5
  are concerned with the effects of mass loss on the baryon surface
  density profile and velocity structure. The * next to run B4 for
  the infall time indicates that this is a special run in which three
  slow inflow phases over 15 time units were punctuated by phases of fast mass
  loss over 1 time unit. The system was evolved for 30 time units
  ($\sim 2$Gyrs) in-between each mass inflow and each mass outflow
  phase.}
\label{tab:initcond}
\end{table*} 

\subsection{Angular momentum}\label{sec:resangmom}

In this section we consider the dynamical effect of mass loss from
two-component galaxies, initially in equilibrium, on the final angular
momentum distribution of the baryons. In doing this we seek to find
out whether or not mass loss can reconcile the observed, low angular
momenta of the Local Group dSph galaxies with the wide range of
angular momenta predicted by cosmological simulations (see section
\ref{sec:introduction}).

We consider first a set of extreme models
where 95\% of the initial baryonic mass is removed on a disc crossing
time - the impulsive mass loss case. These extreme models are relevant
since, if they fail to appreciably perturb the initial angular
momentum distribution, then we can rule out a wide range of more reasonable
mass loss scenarios. 95\% mass loss is also a reasonable place
to start since, if dwarf galaxies formed with the cosmological mean
baryon fraction, then they must have lost $\sim 95$\% of their baryonic
mass by the present epoch (\citet{2003Astro-Ph..0302209} and
\citet{2001ApJ...563L.115K}).

The simulations are labelled, in order of discussion, from A1 through
to A8. The initial conditions and model parameters for all of the runs
are summarised in table \ref{tab:initcond}.

\subsubsection{Quantifying the angular momentum of a remnant
  galaxy}\label{sec:ang:rotfactor}

While in these simulations we are able to measure exactly the angular
momentum distribution of the baryons after mass loss, this is not the
case for real dwarf galaxies observed in projection on the sky. As
such, it is also useful to have a statistical measure of the observable
amount of angular momentum. 

To quantify the measurable amount of angular momentum left in a
remnant system, we performed 500 random realisations of a non-rotating
spheroid with the same velocity dispersion as used for the rotating
spheroid initial conditions, but with only 750 baryon particles to mimic the
effect of a 95\% mass loss. We then define the {\it rotation factor},
$RF$, as: 

\begin{equation}
RF = ||\bar{v_l}-\bar{v_r}|-|\bar{v_t}-\bar{v_b}||
\label{eqn:rotfactor}
\end{equation}

where $\bar{v_i}$ ($i=l,r,t,b$) are the average projected velocities
over the left, right, top and bottom sections of the velocity
projection, rotated so that the maximal rotation axis is vertical and
centred.

For the non-rotating spheroid, the mean rotation factor was 0.01,
while the 3$\sigma$ deviation from the mean was 0.08. We define, then,
a system with {\it statistically significant} rotation as one which has $RF
> 0.08$ - i.e. with a rotation factor which is greater than 3$\sigma$ from
the mean for a non-rotating system. For comparison, the rotation
factors for initial conditions where the baryons were set up in a
rotating disc or rotating spheroid were 0.64 and 0.36 respectively.

\subsubsection{Random mass removal - runs A1-A6}\label{sec:ang:random}

The first case of interest is that where the mass was removed
from randomly selected regions of the baryon spatial distribution. In the
following simulation runs A1-A6, 95\% of the mass of 
the baryons was removed in one baryon crossing time ($\sim 1$ in
simulations units). In runs A1-A3 the baryons were initially set up in
equilibrium with a rotating exponential disc with $M_b=1$,$h_d=1$ and
$z_0=0.2$. A Toomre Q-parameter of 1.5 was used, corresponding to an
initially stable disc \citep{1987gady.book.....B}, although changes to
this parameter were not found to significantly affect the
results. From run A1 through to A3 all that was changed in the initial
conditions was the underlying dark matter profile, with the mass of
dark matter interior to the baryons decreasing from A1 to
A3. Similarly, runs A4-A6 had the same decreasing sequence of dark
matter mass, but the baryons were initially set up in equilibrium with
a rotating Hernquist density profile of mass, $M_b=1$, and scale
length, $h_b=1$. For these runs the initial angular momentum was less
than in the disc cases as discussed in section
\ref{sec:initialcond}. The initial conditions for all of the runs are
summarised in table \ref{tab:initcond}.

\begin{figure*}
\begin{center}
  \epsfig{file=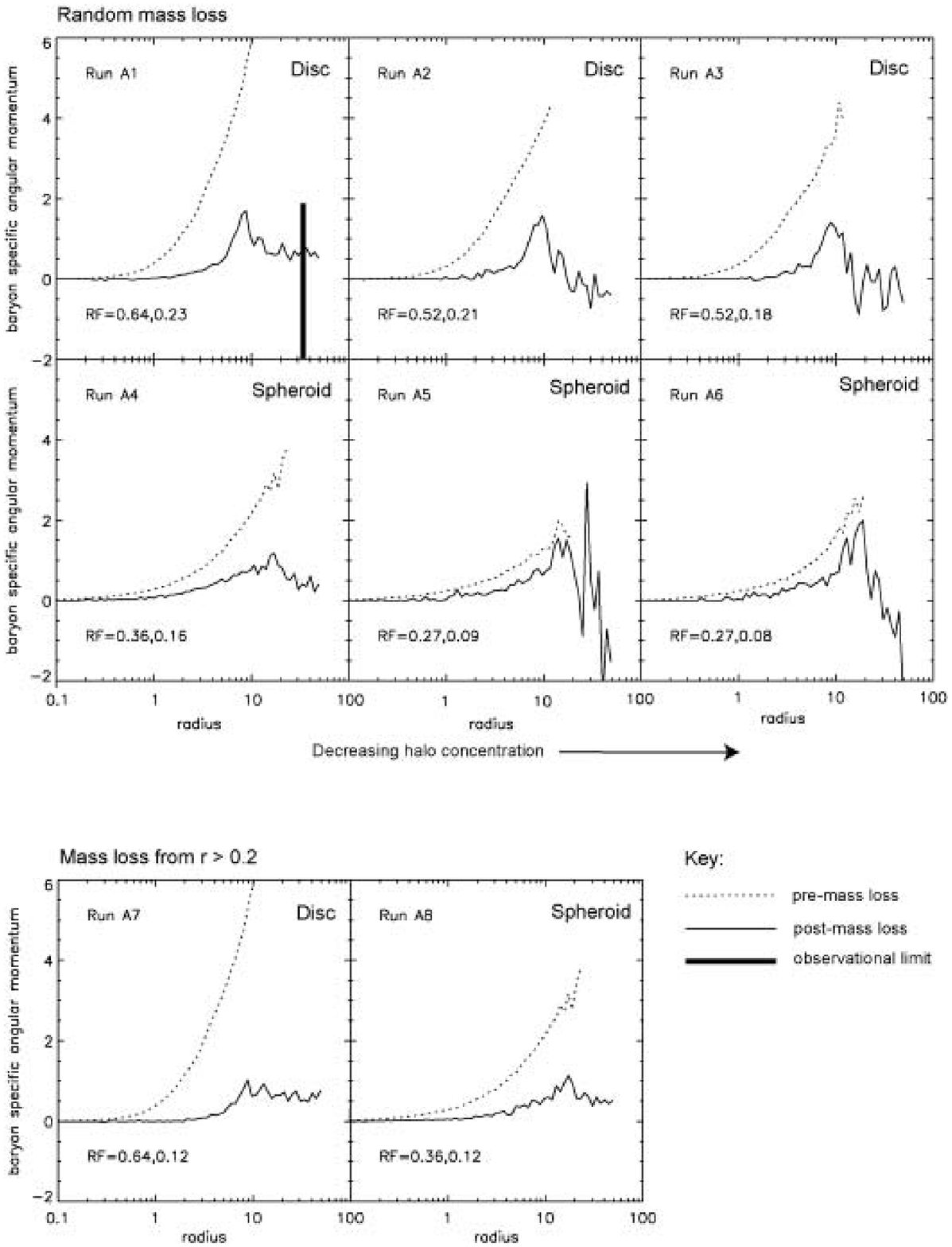,width=120mm}
  \caption[Baryon specific angular momentum profiles before and after mass
  loss for runs A1-A8]
  {Baryon specific angular momentum profiles before (dotted lines) and after
  mass loss (solid lines) for runs A1-A8. The initial
  conditions were as in table \ref{tab:initcond}. The
  units for all of the runs were as in section \ref{sec:units}. Marked
  in the bottom left corner of each run is the rotation factor (see
  equation \ref{eqn:rotfactor}) before (left of comma) and after
  (right of comma) the mass loss. A rotation factor of $>0.08$ represents
  statistically significant rotation at the $>3\sigma$ level. The top
  six plots (runs A1-A6) show the effect of random mass loss for
  different initial conditions. From left to right, the dark matter halo
  concentration is reduced for disc (top) and spheroid
  (bottom) initial conditions. The bottom two plots (runs A7 and A8)
  show the effect of removing the mass from all radii larger than 0.2
  baryon scale lengths for disc (left) and spheroid (right) initial
  conditions. Marked on run A1 is the current observational limit,
  which is the same for all of the runs, 
  where the surface brightness drops below the background noise.
  Notice that this limit lies beyond the
  peak in the specific angular momentum after mass loss in all cases.}
\label{fig:angmomplot}
\end{center} 
\end{figure*}

Figure \ref{fig:angmomplot} displays the specific angular
momentum of the baryons before (dotted lines) and after mass loss
(solid lines) for runs A1-A8. Notice that, as the initial dark matter
mass interior to the baryons decreases from run A1 to A3, the decrease in
angular momentum is greater. The same trend is observed in runs A4 to
A6.

The rotation factors, marked in the bottom left corner of
each plot, follow the same trend as
discussed above for the baryon specific angular momentum after mass
loss. Only one run (A6)
falls below statistically significant rotation at the 3$\sigma$ level
after mass loss. This run represents the most extreme model where the
baryons are initially, largely pressure supported ($v_r/\sigma =
0.8$\footnotemark) and where the initial dark matter mass interior to
the baryons is very low ($\sim 1$\% of the initial baryon mass).
\footnotetext{In this context $v_r$ is the rotational velocity of the
  baryons at the edge of the light distribution, while $\sigma$ is the
  mean velocity dispersion averaged over the whole distribution. $v_r$
  is averaged over 
  a radial slice of thickness 0.5 in simulation units and at an outer
  radius, $r_l$, where $v_r$ is maximised. We do not allow the
  particle number at $r_l$ to fall below order 10 so as to avoid
  sampling at large radii where statistical noise from low particle
  numbers can lead to spurious results. For the non-rotating spheroid
  discussed in \ref{sec:ang:rotfactor}, $v_r/\sigma = 0.04$.}

\subsubsection{Biased mass removal - runs A6-A8}\label{sec:ang:biased}

While the spatially-random mass removal cases do reduce the angular
momentum of the progenitor galaxy, in nearly all of the above
scenarios significant angular momentum remains in the baryons after
mass loss. However, for 
both the disc and spheroid initial conditions, the majority of the
angular momentum is contained at large radii (see figure
\ref{fig:angmomplot}). While not linked to a specific physical model
of dSph star-formation and 
feedback, it is interesting to consider the case of biased mass loss,
where mass is lost preferentially from the outer parts of the disc or
spheroid. In this way, we preferentially remove the high angular
momentum material. This scenario could be of importance for models
where material (stellar, gas and dark matter) is
tidally stripped from the outer regions through interactions with a
massive host galaxy (see e.g. \citet{2001ApJ...547L.123M},
\citet{2002MNRAS.335..487M}, \citet{1998ARA&A..36..435M} and section
\ref{sec:introduction}).

Figure \ref{fig:angmomplot} runs A7 and A8 shows the effect of the
removal of all of the mass from the disc and spheroid respectively at
radii larger than the disc scale height. This amounts
to a mass loss of 99\% for the disc (run A7) and 97\% for the spheroid
(run A8) and should remove all the high angular momentum parts, while
leaving a small, approximately spherical, remnant behind.

As can be seen, this biased mass
removal has largely worked as one would expect. Much of the angular
momentum of the baryons has been removed, while the 
rotation factors for the remnants (as defined in section
\ref{sec:ang:rotfactor}) are reduced to 0.12 for both the disc and spheroid
cases. However, even in this extreme case, in both runs A7 and A8
statistically significant rotation remains in the remnant galaxies
after mass loss. The relevance of these results will be discussed
further in section \ref{sec:observations}.

\subsubsection{Asymmetric mass loss}\label{sec:ang:asymmetric}

All of the mass loss scenarios considered above far have been
spherically symmetric. For tidal stripping, however, or other possible
mechanisms of mass loss linked to local star formation rates, the
assumption of spherical symmetry 
could be very poor. As such, we also considered the effects of
asymmetric mass loss. The results from these runs were found not to
significantly differ in terms of angular momentum loss from those
presented above: angular momentum is still difficult to remove through
mass loss.

\begin{figure*} 
\begin{center}
  \epsfig{file=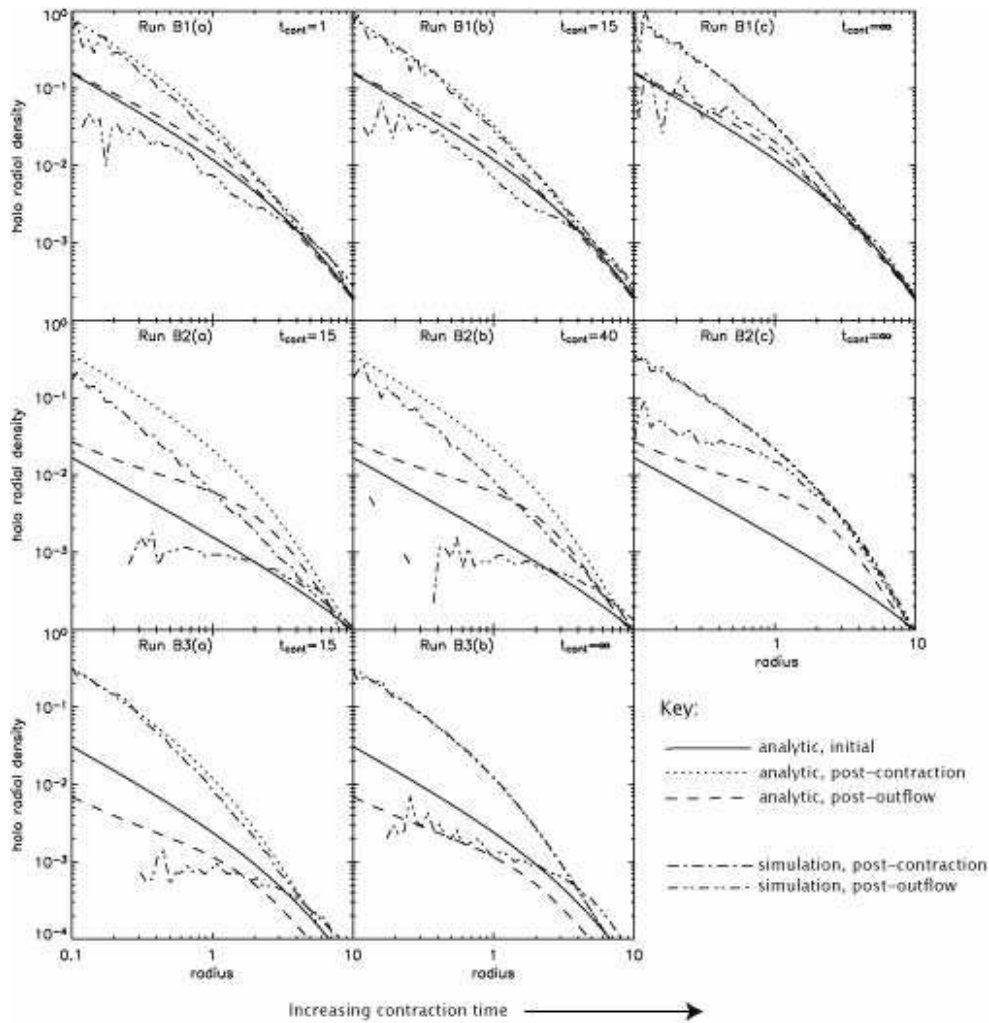,width=130mm}
  \caption[Runs B1-B3: Response of a centrally cusped dark matter halo to
  the slow addition and fast removal of some baryons]
  {Runs B1-B3: Response of a centrally cusped dark matter halo to the
  slow addition and fast removal of some baryons. The units for the
  plot are as in section 
  \ref{sec:units}. The initial conditions for each run were as in table
  \ref{tab:initcond}. Run B1 had initial conditions close to the
  cosmological mean. Runs B2 and B3 were identical to run B1 but with
  long scale length and low mass dark matter halos respectively. The
  baryon infall time, $t_{cont}$, for each run 
  is marked in the top right corner and is increasing from left to
  right. Where $t_{cont}=\infty$, the
  contraction of the halo due to the slow addition of the baryons was
  calculated analytically as in appendix \ref{sec:zhaocalc}. 
  The dot-dashed lines show the halo density after the
  baryons have been added to the system over a time, $t_{cont}$, while
  the double-dot dashed lines show the halo density after a subsequent
  impulsive mass loss over 1 time unit ($\sim 1$ baryon crossing
  time). The solid lines show the initial halo density profile,
  the dotted lines show the analytic prediction for the halo
  density profile for adiabatic mass inflow and the dashed lines
  show the heuristic analytic prediction for the final halo density
  profile after mass loss as derived in appendix
  \ref{sec:zhaocalc}. All of the runs were evolved for 15 time units
  ($\sim1$Gyr) after mass loss. Notice that as the 
  inflow time is increased, the contraction of the dark matter halo
  due to the addition of the baryons (dot-dashed lines) tends towards
  the adiabatic limit (dotted line). Notice, further that, even
  when the non-adiabatic inflow cases ($t_{cont} \neq \infty$) produce
  contracted density profiles which are very close to the adiabatic
  limit, the resulting density profiles after mass loss are much
  shallower than the true adiabatic case.}
\label{fig:efdm:denplate}
\end{center}    
\end{figure*}

\subsection{The dark matter density profile}\label{sec:resdarkmatter}

In this section we consider the dynamical effect of mass loss from
two-component galaxies, initially in equilibrium, on the underlying
dark matter distribution. Recall from section \ref{sec:introduction}
that the UMi dSph galaxy appears to have a near-constant central dark matter
density profile, whereas cosmological simulations predict a central log
slope for the density profile of $1 < \alpha < 1.5$. Here we consider
whether or not mass loss could explain this discrepancy.

For brevity we present the results from five simulations involving
different extrema initial conditions, labelled B1 through to B4. A
summary of the initial conditions and model parameters for all of the
runs is displayed in table \ref{tab:initcond}.

Overlaid on the simulation results are also analytic calculations for
the adiabatic response of the halo to baryon dissipation and collapse
and the subsequent response of the halo to an impulsive mass loss from
the baryons. These calculations are very similar to those presented in
\citet{2002MNRAS.333..299G} and are outlined for completeness in
appendix \ref{sec:zhaocalc}. The analytic calculations are useful
for two reasons: they provide an independent check that the numerical
code is producing sensible results, and they highlight the important
physics involved. For slow gas dissipation and infall, angular
momentum, mass and orbital structure must be conserved (i.e. those
particles initially on circular orbits will remain on circular
orbits). For fast impulsive outflow, or faster gas inflow, angular
momentum and mass are still conserved (provided the infall is
spherically symmetric) but an initially circular orbit will become a
more energetic (more radial) orbit. It is this structural change which
leads to an altered density profile in the dark matter after mass
loss.

\subsubsection{The choice of inflow time}\label{sec:inflowtime}

For all of the runs presented in this section we allowed the
dark matter halo to contract in response to the addition of the
baryons, either by slowing increasing the mass of the baryons over a
time, $t_{cont}$, while holding the baryons fixed; or by using an
analytic prescription for adiabatic mass infall (see appendix
\ref{sec:zhaocalc}). We then allowed the system to evolve
for a further 30 time units before any mass was removed; the results
were found not to be sensitive to this parameter.

To see why it is important to consider a range of inflow timescales
for the baryons, it is worth considering the free fall time, $t_{ff}$,
for a pressureless gas sphere of uniform density,
$\overline{\rho}$. This is given by \citep{1987gady.book.....B}:

\begin{equation}
t_{ff} = \sqrt{\frac{3\pi}{32G\overline{\rho}}}
\label{eqn:efdm:tff}
\end{equation}
\noindent
In simulation units, for typical initial conditions with $M_h=a_h=10$,
this gives $t_{ff} \sim 11$. 

We can gain a crude handle on how close to the adiabatic regime the
gas infall is by comparing this gas inflow time with the crossing time
for a dark matter particle at the baryon scale length,
$t_{cross}(h_b)$. This gives \citep{1987gady.book.....B}:

\begin{equation}
t_{cross}(h_b) = \sqrt{\frac{h_b}{G}\left(\frac{M_h}{(h_b+a_h)^2}+\frac{M_b}{4
      h_b^2}\right)^{-1}}
\label{eqn:efdm:tcrossastrue}
\end{equation}
\noindent
During the mass inflow phase, we can then find an {\it upper bound}
for $t_{cross}$ by assuming that $M_b=0$ throughout the contraction.

For $t_{cont} \gg t_{cross}(h_b)$, we should achieve close to
adiabatic behaviour. Notice that for $M_h=10$ and $a_h=10$,
$t_{cross}(h_b)\sim 4$ units. Thus, even gas inflow on a free fall
timescale may be non-adiabatic. Faster gas infall due to, for
example, mergers \citep{2001cksa.conf..735N} or interactions
\citep{1994ApJ...425L..13M} will then be even further away from the
adiabatic limit, and so it is important to consider the case of
non-adiabatic inflow.

For most of the runs, we will use $t_{cont}=15$ units, which
translates to $t_{cont} \sim 0.7 \times 10^9$years - of order a
giga-year, for a $10^8$M$_\odot$ galaxy with scale length 1kpc (see
section \ref{sec:units}).

\subsubsection{The effect of varying the inflow time - run
  B1}\label{sec:efdm:results}

Run B1 (see figure \ref{fig:efdm:denplate}) had initial conditions set
up as in table \ref{tab:initcond}:
$M_h=10, a_h=10, t_{ml}=1$ and $\delta=0.95$. This gave a baryon
fraction, $f_b = \frac{M_b}{M_b+M_h} = 0.1$ initially and $f_b=200$
after the mass loss. The three plots B1(a), B1(b) and 
B1(c) show the effect of varying the baryon infall time, $t_{cont}$,
which is marked on each. For plot B1(c), where $t_{cont}=\infty$, the
contraction of the halo due to the slow addition of the baryons was
calculated analytically as in appendix \ref{sec:zhaocalc}. For all
three plots, the dot-dashed lines show the halo density after the
baryons have been added to the system over a time, $t_{cont}$, while the
double-dot dashed lines show the halo density after a subsequent
impulsive mass loss 
over 1 time unit ($\sim 1$ baryon crossing time). The solid line
shows the initial halo density profile, the dotted line shows the
analytic prediction for the halo density profile after adiabatic mass
inflow and the dashed line shows the heuristic analytic
prediction for the final halo density profile after mass loss as
derived in appendix \ref{sec:zhaocalc}. Notice that there is a small
core interior to $\sim 0.2$ baryon scale lengths due to 
particle noise and two-body relaxation (see section
\ref{sec:eqtests}) which can be seen in all of the simulations. This
effectively sets the simulation resolution. 

Although in all of the runs the
contracted halo density profile seems very close to the adiabatic
limit (see the double-dot dashed line as compared with the dotted line),
the post-mass loss density profiles are quite different (compare the
dot-dashed lines in plots B1(a)-B1(c)). For the true adiabatic run, B1(c),
the final halo density profile after mass loss is in reasonable
agreement with the heuristic 
analytic prediction from appendix \ref{sec:zhaocalc} (see double-dot
dashed lines as compared 
with the dashed line). However, for the faster inflow time runs,
the final density profile after mass loss is much shallower. The
reason for this is that
non-adiabatic inflow (that is inflow on a timescale which is short as
compared with the halo dynamical time), does not preserve the initial
orbital structure. The initially isotropic halo becomes strongly radially
anisotropic after mass inflow. This anisotropy then significantly increases
the number of halo particles which become unbound after 
subsequent mass loss and leads to much shallower final halo density
profiles (see appendix \ref{sec:zhaocalc}). 

\begin{figure} 
\begin{center}
  \epsfig{file=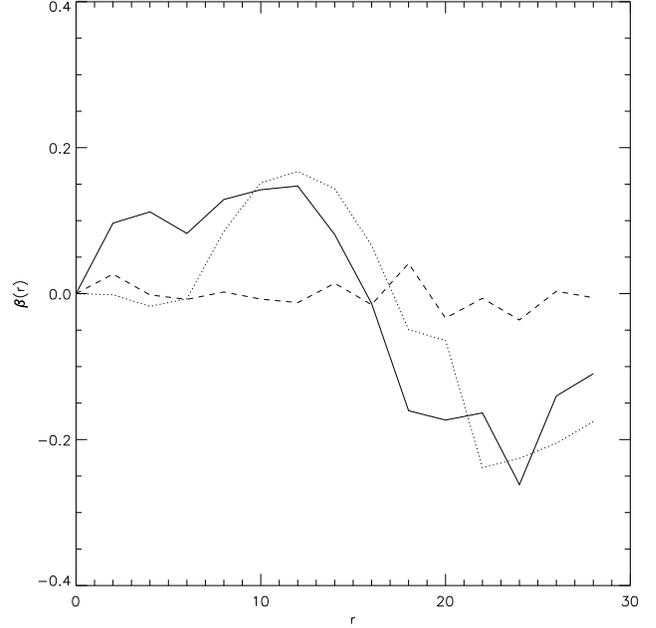,width=0.5\textwidth}
  \caption[The effect of the baryon inflow time on the halo velocity
  anisotropy]
  {The effect of the baryon inflow time on the halo velocity
  anisotropy. The lines in the plot show the radially binned halo
  velocity anisotropy, $\beta(r)$, after mass inflow for run B1. The
  solid line is for inflow on 1 baryon crossing time (run B1(a)), the
  dotted line is for in flow over 15 baryon crossing times (run B1(b))
  and the dashed line is for the infinitely slow limit of adiabatic
  inflow, which was set up analytically as in appendix
  \ref{sec:zhaocalc} (run B1(c)). Notice that the
  amount of anisotropy introduced {\it increases} as the infall time
  is {\it reduced}.}
\label{fig:efdm:betacompare}
\end{center}    
\end{figure}

This is further illustrated by figure \ref{fig:efdm:betacompare},
which compares the halo velocity anisotropy, $\beta(r)$, after mass
inflow for the different simulation runs B1(a) to B1(c)
($\beta(r)=1-\overline{v_\theta^2}/\overline{v_r^2}$ is defined as in
\citet{1987gady.book.....B}). The solid line is for run B1(a), the
dotted line is for B1(b) and the dashed line is for B1(c). Notice that
the dashed line for run B1(c) has $\overline{\beta(r)}=0$, as expected for
initial conditions which are set to be isotropic. For the other runs,
however, where the cusp contraction was performed numerically, some
anisotropy has been introduced to the system. The
amount of anisotropy introduced {\it increases} as the infall time is
{\it reduced}; notice that the solid black line in figure
\ref{fig:efdm:betacompare} displays more anisotropy than the dotted
line - particularly at small radii. Finally, notice that $\beta(r)$
actually becomes negative at $\sim$ the halo scale length. This is
because while the initially isotropic dark matter distribution will have
$\overline{\beta(r)}=0$, some particles will be more radially
anisotropic than others. After mass inflow, these particles will be
more efficiently heated and pushed to larger radii. For mass inflow
which produces a small anisotropy, as in this case, this can lead to a
depletion in radially anisotropic particles at $\sim a_h$ and hence a
negative value for $\beta(\sim a_h)$.

\subsubsection{The effect of increasing the collapse factor - run B2}

Run B2 (see figure \ref{fig:efdm:denplate}) demonstrates the effect of
increasing the collapse factor\footnote{The
  nomenclature - collapse factor - comes from the assumption that the
  baryons initially have the same scale length as the dark matter
  ($a_h$) and then collapse (cool) to form a smaller rotationally supported
  or pressure supported system with scale length $h_b$ (see appendix
  \ref{sec:zhaocalc}).}, defined as $a_h/h_b$ (the ratio of the
halo and baryon scale lengths). The initial
conditions were as in run B1 but with $a_h=50$ and hence with collapse factor
$a_h/h_b=50$. The data points and curves are as for run B1, with the
contraction times, $t_{cont}$ similarly marked. Table
\ref{tab:initcond} shows that these runs required much higher 
numerical resolution, with 681,818 halo particles. This is because
longer scale length halos put more of the mass (and hence the
particles) out at large radii. Thus, to achieve the same local resolution as
in run B1, we must increase the number of particles at the centre by
increasing the total number of particles. As in run B1,
the effect of changing the baryon infall time from the adiabatic
case (plot B2(c)) to the non-adiabatic cases (plots B2(a) and B2(b))
produces shallower final density profiles. The effect, however,
is magnified as compared with run B1. This is because the crossing time for 
the halo at $h_b$ (now with a much longer scale length) has significantly
increased (see equation \ref{eqn:efdm:tcrossastrue}) from $\sim 4$ units
for run B1 to $16$ units for run B2. Thus, in run B2(a), a
contraction time of $t_{cont}=15$ is much further away
from adiabaticity than the same contraction time is in run
B1(b): the effect is now large enough to produce significantly
shallower density profiles after mass inflow too. As might be expected
though, the longer inflow time run B2(b) is closer to adiabaticity than
the short inflow time run B2(a) (compare the
dot dashed and dotted lines for both of these runs).

The final density profiles after mass loss - even
for an initial cusp contraction which is adiabatic - do not agree well with
the heuristic, analytic, prediction presented in appendix
\ref{sec:zhaocalc}. This is, 
perhaps, not surprising since the analytic 
formula was derived by considering only the limiting cases of
impulsive mass loss and adiabatic mass loss. For the impulsive case,
the limiting behaviour was taken such that the system would become
unbound if it suddenly lost half of its mass. For the extreme scenario
presented in run B2, where there is relatively little central dark matter
mass initially, this heuristic argument clearly breaks down.

For these extreme initial conditions, a core does
form in all cases in the final halo density profile.

\subsubsection{The effect of increasing the baryon fraction - run B3}

Run B3 (see figure \ref{fig:efdm:denplate}) demonstrates
the effect of increasing the baryon fraction. The 
initial conditions were as in run B1 but with $M_h=2$ and so
$f_b=0.33$. Now, as in run B2, a central core with scale length $\sim
1-2$ forms in the halo density profile. There is 
little difference in the contracted density profiles (dot-dashed lines)
between the adiabatic (plot B3(b)) and non-adiabatic (plot B3(a)) infall
cases, yet the crossing time at $h_b$ for the halo is 8.4 units;
dropping the mass in over 15 time units is clearly {\it not} a good
approximation to the true adiabatic case. As with run B1, this is why,
after mass loss, the density profile for the non-adiabatic inflow run
B3(a) is shallower than that for the adiabatic case B3(b).

The adiabatic case, as in run B1, is reasonably well fit by the heuristic
calculation for impulsive mass loss presented in appendix
\ref{sec:zhaocalc}. As with
run B2, a core has formed in the final density profile after mass 
loss for the non-adiabatic infall (plot B3(a)). The similarity between
runs B2 and B3 after mass loss should not be too surprising.
The factor which really matters - the central mass of dark matter as
compared with the initial central baryonic mass - is very low in both
cases. For run B2 this is achieved by pushing most of the dark matter
out to large radii; for run B3 it is achieved by reducing the total
halo mass.

\begin{figure} 
\begin{center}
  \epsfig{file=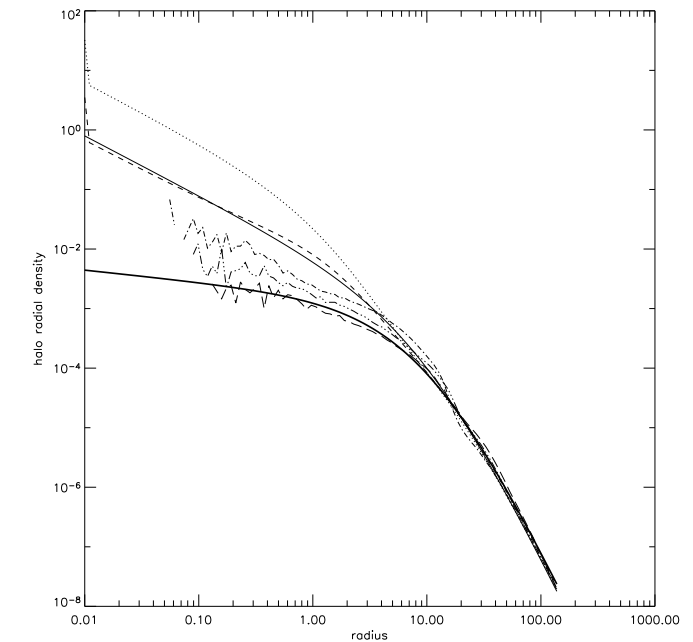,width=0.5\textwidth}
  \caption[The effect of multiple inflows and outflows on the central
  halo density profile]
  {The effect of multiple inflows and outflows on the central
  halo density profile. The initial conditions for this run were as in
  table \ref{tab:initcond}, while the units were as in section
  \ref{sec:units} and were for the cosmological mean values. In this run the
  baryons were dropped in slowly ($t_{cont}=15$) and 95\% removed
  quickly ($t_{ml}=1$), {\it three times}. The resulting halo density
  profile after one two and three mass loss phases is marked by the
  dot-dashed, triple-dot dashed and long-dashed lines
  respectively. The thick solid line on this plot shows a density
  profile with central log-slope of $\alpha=0.2$ for comparison with
  the simulation data. All of the other lines are as in figure
  \ref{fig:efdm:denplate}.}
\label{fig:efdm:runb4}
\end{center}    
\end{figure}

\subsubsection{The effect of multiple mass inflows and outflows - run
  B4}\label{sec:efdm:multiout}

We have so far considered only the case of one slow mass infall phase
followed by one impulsive mass loss phase. However, this process may
have happened in dSph galaxies more than once, either at very early
times or perhaps for some dSph throughout their history (the Carina,
Leo I and Sagittarius dSphs all show evidence for continuous episodic
star formation over the past $\sim$14Gyrs (\citet{2003AJ....126..218M},
\citet{2000MNRAS.317..831H} and \citet{2002MNRAS.332...91D}). As
such, it is interesting to quantify the dynamical effect of such continuing
mass inflow and outflow on a dSph galaxy on the underlying dark matter
density profile.

Run B4 (see figure \ref{fig:efdm:runb4}) demonstrates the effect of
multiple mass loss phases punctuated by periods of slow accretion. The
initial conditions were as in table \ref{tab:initcond} and represent the
cosmological mean, with $f_b = 0.17$
\citep{2003Astro-Ph..0302209}. In this run the baryons were dropped in
slowly ($t_{cont}=15$) and 95\% removed quickly ($t_{ml}=1$), {\it
  three times}. The resulting halo density profile after one two and
three mass loss phases is marked by the dot-dashed, triple-dot dashed
and long-dashed lines respectively. The thick solid line on this plot
shows a density profile with central log-slope of $\alpha=0.2$ for
comparison with the simulation data. All of the other lines are as in
figure \ref{fig:efdm:denplate}.

Notice that after one mass inflow and outflow phase, as with run B1,
the central dark matter density after mass loss is shallower than in
the initial conditions (see the dot-dashed line compared with the
solid line). It is this asymmetry which, when iterated three times,
produces the much shallower central density profile in figure
\ref{fig:efdm:runb4} (long-dashed line). The central density profile
after three mass loss events is reasonably well fit by a
shallower central dark matter slope
with $\alpha \sim 0.2$ (see the thick solid line). Even after two such
mass loss events, the central dark matter density 
profile has a central core (see triple-dot-dashed line). The
`core size' for the final density profile in figure
\ref{fig:efdm:runb4}, for both two 
and three mass loss events is $\sim 1-2$ baryon scale lengths. This
translates to a $\sim 1$kpc core for a dSph at the present epoch with
scale length 1kpc.

\subsection{The baryon surface density profile}\label{sec:ressurfden}

In this section we consider the dynamical effect of mass loss from
two-component galaxies, initially in equilibrium, on the surface
density profile and 
projected velocity dispersion of the baryons. It is important to
quantify the effect of mass loss on these observational signatures
since these can then be compared with measurements from dwarf galaxies
in the Local Group (see section \ref{sec:observations}).

The initial conditions used in this section were for a generalised
Hernquist spheroid for the baryons (see equation \ref{eqn:spheroid})
and a Hernquist profile for the dark matter (see equation
\ref{eqn:hernquist}). The generalised Hernquist profile was useful for
the baryons as it allows for an easy parameterisation of a wide range
of initial surface density profiles. The baryons were set up initially
to be fully pressure supported with no internal angular momentum. This
choice was motivated by the results from section \ref{sec:resangmom}
which suggest that spatially random mass loss produces a small
perturbation on the total internal angular momentum in dSph galaxies,
and by observations from the Local Group which suggest that dSph
galaxies at the present epoch have very little internal angular
momentum (see section \ref{sec:introduction}). However, when rotating
initial conditions were used, similar results to the non-rotating
cases were obtained. As such, we present only the angular momentum
free models here.

Similarly, while many other runs were also performed to
demonstrate that the results are not sensitive to the choice of
initial baryon or dark matter density profiles, for brevity we present
only the key runs which illustrate the main points of interest.

The simulations are labelled, in order of discussion, from C1 to
C5. Unless otherwise stated, all of the runs had the same initial
conditions as run C1 but vary the outflow time and/or allow for baryon
dissipation and collapse as in section \ref{sec:resdarkmatter} before
mass loss. The initial conditions for all of the runs are summarised
in table \ref{tab:initcond}.

Overlaid on the simulation results are also analytic calculations for
the response of the baryons to an impulsive gas mass loss detailed in
appendix \ref{sec:surfdencalc}. As with section
\ref{sec:resdarkmatter}, the analytic calculations provide a useful
independent check of the numerical code while also highlighting the
important physics.

\subsubsection{Impulsive outflow - runs C1 and C2}

\begin{figure*} 
\begin{center}
  \epsfig{file=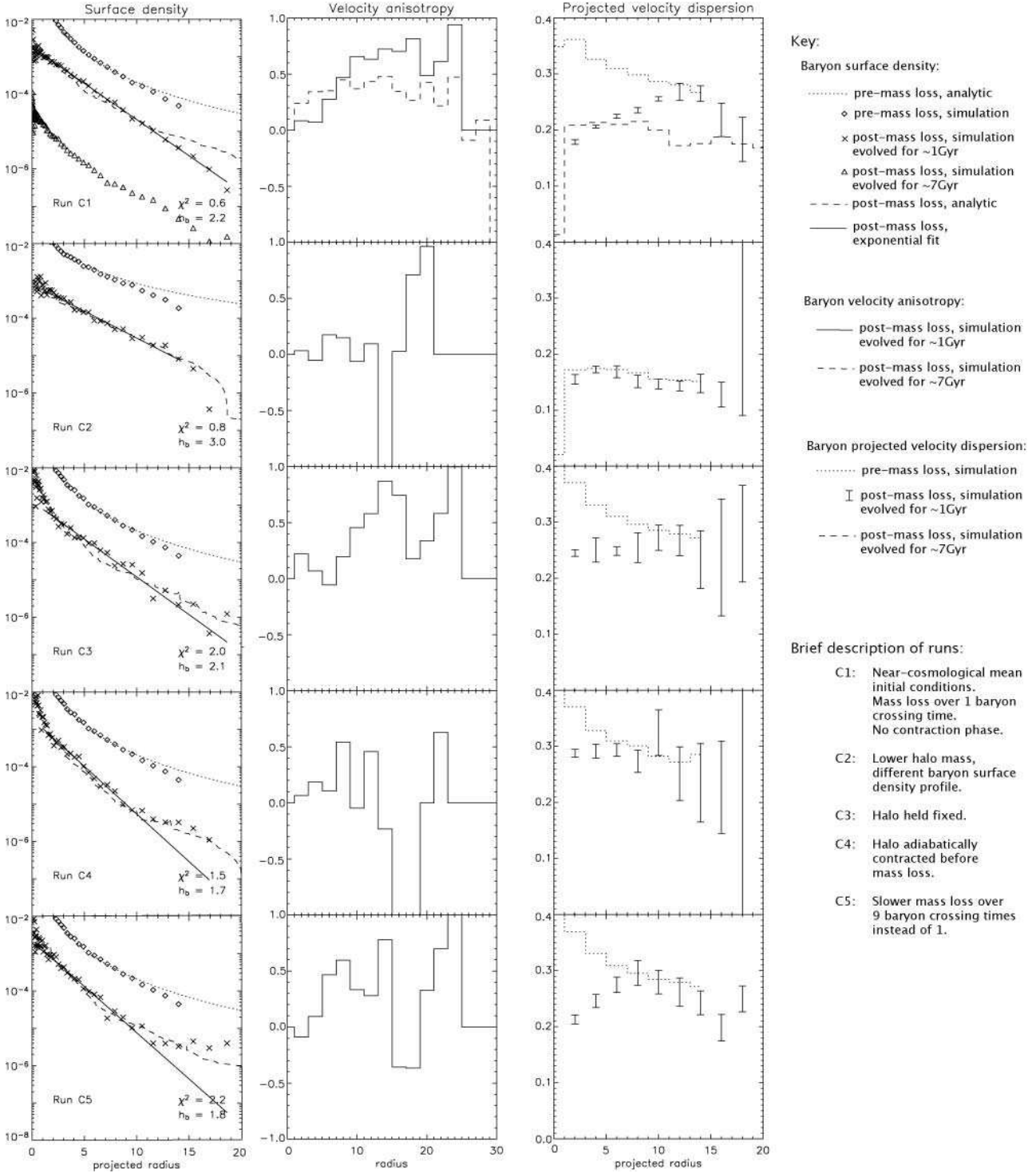,width=170mm,height=195mm}
  \caption[]
  {The response of the baryons to mass loss from the baryons, runs
  C1-C5. The left panel shows 
  the baryon surface density profile integrated along the z-axis
  before (diamonds) and after (crosses) the mass loss. The dotted
  lines are the initial analytic surface density profiles, the 
  dashed lines are the analytic predictions for the final density
  profile after mass loss in the impulsive limit. The straight solid
  line is an exponential fit to the final baryon surface density    
  profile (crosses). $\chi^2$ for the fit is
  shown in the bottom right of the surface density plot, along with the
  fitted baryon scale length: $h_b$. The middle panel shows the
  baryon velocity anisotropy, $\beta(r)$, as a function of radius
  after the mass loss. The
  right panel shows the projected velocity dispersion as a function of
  projected radius for the baryons before mass loss (dotted lines) and
  after mass loss (data points), averaged over 20 different
  projection angles. The error bars are determined from the standard
  deviation between results obtained from different projection
  axes. All of the runs were evolved for 15 time units ($\sim1$Gyr)
  after mass loss except for run C1 which was also 
  evolved for a further 100 time units (see the triangles in the left
  panel and the dashed lines in the middle and right panels) to test for
  equilibrium. The triangles have been offset from their true value by
  $10^{-2}$ for clarity. The initial conditions for the runs are as in
  table \ref{tab:initcond}, while the units are as in section
  \ref{sec:units}. The initial conditions for run C1 were a Hernquist,
  non rotating spheroid of stars with $M_b=h_b=$1, embedded in a
  Hernquist halo with $M_h=a_h=$10. The rest of the runs consider the
  effect of altering either these initial conditions or the mass loss
  rate as detailed in the text.}
\label{fig:sfdn:surfden}
\end{center}    
\end{figure*}

Figure \ref{fig:sfdn:surfden}, top two rows, show the results for runs
C1 and C2. The purpose of these simulations was to see
whether, for fast mass loss from a galaxy in which the baryons
initially dominate the central potential, the baryons retain
information about their initial density profile and velocity
dispersion. 

In these runs, the halo was not adiabatically contracted to account for
the presence of the baryons. This allowed the initial conditions to be
carefully controlled: we ensured that the velocity distribution before
mass loss for both the baryons and the dark matter was isotropic. 

After mass loss, the surface density profile of the 
baryons in both runs C1 and C2 (crosses, left panel)
are now well fit by an exponential (solid straight line), irrespective of
the initial conditions.

The dashed line in the left panel of figure \ref{fig:sfdn:surfden}
shows the post-mass loss baryon 
surface density profile calculated analytically for impulsive mass
loss as in appendix \ref{sec:surfdencalc}. As can be seen, this
analytic calculation provides a good fit to the simulation data over
many baryon scale lengths. This good agreement highlights two
important points: firstly, that the numerical code is producing
sensible results; and secondly, that the mechanism driving the
change in surface density profile is the change in orbital structure
of the baryons produced by the fast mass loss (see appendix
\ref{sec:surfdencalc}). For fast changes to the local potential,
orbital structure is no longer conserved - all orbits become more
energetic and, therefore, more radial (since angular momentum will
still be conserved for symmetric changes). The time integration of these
orbits leads to an approximately exponential surface density profile
over many radii after mass loss (see appendix \ref{sec:surfdencalc}).

The middle panels of figure \ref{fig:sfdn:surfden} show the baryon velocity
anisotropy, $\beta(r)$,
after mass loss. The first thing to note is that the baryons, which
were initially isotropic before mass loss with $\beta(r)=0$, now show
significant radial anisotropy as a result of the mass loss. The second
thing to note is that in all cases there is a drop in $\beta(r)$ at
$\sim$ the halo scale length ($a_h$=10 units in figure
\ref{fig:sfdn:surfden}). After mass loss, at $\sim a_h$,
the baryons move from being radially anisotropic as a result of the
mass loss to being closer to their initial conditions with $\beta(r)
\sim 0$. This is because the halo scale length sets the scale at which
the changes to the central potential caused by baryon mass loss become
negligible.

For two of the simulations - runs C2 and C4 - $\beta(r)$ actually
  becomes {\it 
  negative} at $\sim a_h$ after mass loss. This occurs for similar
reasons to $\beta(r)$ becoming negative at large $r$ for the dark
matter halo in figure \ref{fig:efdm:betacompare}: the more radially
anisotropic baryon particles before mass loss are preferentially moved
towards larger radii after mass loss. For the simulations where the
anisotropy introduced by the mass loss is small (as is the case for
runs C2 and C4), this can deplete the fraction of radially 
anisotropic particles at $\sim$ the halo scale length after mass loss,
leading to a negative $\beta(r)$. 

After the drop point at $\sim a_h$, $\beta(r)$ rises steeply towards
unity - indicating highly radial orbits. This is because at these
large radii, there are now {\it no particles left from the original
  baryon distribution before mass loss}. The particles which populate
these radii after mass loss are those on highly radial orbits which have
moved outwards as a result of the mass loss.

The right panels of figure \ref{fig:sfdn:surfden} show the projected baryon
velocity dispersion, $\sqrt{\overline{v_p^2(R)}}$, before mass loss (dotted
lines) and after mass loss 
(data points). There are three important features to notice:\\
\\
1. $\sqrt{\overline{v_p^2(R)}}$ is significantly reduced at radii interior to
$\sim a_h$. This is because particles which before mass loss had high
kinetic energy, become easily unbound after mass loss and move outward to
larger radii.\\
\\
2. As with $\beta(r)$, at $\sim a_h$ where the halo
begins to dominate the local potential, the 
projected velocity dispersions converge on the initial condition
values.\\
\\
3. At very large radii, in run C1 there is a
sharp drop in the projected velocity dispersion {\it beyond the
  edge of the initial baryon distribution}. As discussed above, this
is caused by these radii being populated solely by particles on highly
radial orbits.\\
\\
Finally, run C1 was also evolved for a further 100 time units ($\sim
7$Gyrs) after mass loss to test for equilibrium (see triangles and
dashed lines in the first row of figure \ref{fig:sfdn:surfden}). While the halo
was found to evolve very little over this time, the baryons are still
evolving since particles which became unbound after mass loss take
some time to leave the system. Recall from
section \ref{sec:massloss}, that we have not studied in general the
longer term evolution of these dynamical models since it is likely
that over these longer times external tidal effects become important and should
be included in the analysis. However, notice that over longer times, the
baryons continue to slowly expand, with the radial anisotropy
spreading to smaller radii as highly radial particles return from
larger radii. There are now two pronounced distributions: interior to
the original particle distribution, the profile is still approximately
exponential; exterior to the original distribution, the particles are
more radial, causing a drop in the projected velocity dispersion and a
break in the surface density profile at $\sim a_h$.

\subsubsection{Impulsive outflow in a rigid halo potential - run C3}

Figure \ref{fig:sfdn:surfden}, row 3, shows the results for run
C3, which was the same as run C1 but with the halo held fixed,
forming a rigid potential. The purpose of this run was to test whether
the formation of exponential surface density profiles
depends only on the mass loss rate and not the response of the dark
matter halo. As can be seen from figure \ref{fig:sfdn:surfden}, 
with the halo held fixed, the final surface density profile of the
baryons is still reasonably well fit ($\chi^2 = 2$) by an
exponential. However, the fit is much poorer than in run C1. The
baryons retain much more information about their initial condition
density profile. This suggests that the reduced effect of the mass
loss (caused by holding the halo fixed) leads to a smaller
rearrangement of the baryon surface density profile.

The velocity structure for run C3 is qualitatively similar to that in run
C1 (see figure \ref{fig:sfdn:surfden}): the mass loss causes radial
anisotropy to be introduced to 
the remaining baryons; a sharp drop in $\beta(\sim a_h)$ and a
subsequent rise leads to a drop in the projected velocity
dispersion at $\sim a_h$.

\subsubsection{Slow inflow, Impulsive outflow - run C4}

Figure \ref{fig:sfdn:surfden}, row 4, shows the results for the run
C4, in which the baryons were adiabatically contracted using the
prescription set out in appendix \ref{sec:zhaocalc}, after which
95\% of them were removed quickly in 1 time unit. The initial
conditions for this run are otherwise identical to those for run C1.

This run is interesting as it approximates the case where the baryons
collapse dissipatively on some timescale before mass loss.

As with the fixed halo model, run C3, the post-mass loss baryon
surface density distribution in runs C4 is reasonably well fit by an
exponential profile over many radii with $\chi^2 = 1.5$. However, as
with run C3, the baryons still retain more memory of their initial
conditions than in run C1. Where the effect of the mass loss is
reduced, in this case due to the halo contraction before mass loss,
the final baryon surface density profile is closer to the initial
conditions.

The velocity structure is qualitatively similar to run C2 (see figure
\ref{fig:sfdn:surfden}). This again
reflects the fact 
that the baryonic mass loss has a smaller affect on the central
potential in run C4 as compared with run C1.

\subsubsection{Varying the outflow rate - runs C5}

Figure \ref{fig:sfdn:surfden}, row 5, shows the results for the run
C5, in which 95\% of the baryons were removed over a longer
outflow time of $t_{ml}=9$. The purpose of this
simulation was to quantify how fast the mass loss must be in order to produce
a surface density profile which is well fit by an exponential. The
initial conditions for this run were in all other respects identical
to run C1.

As can be seen from figure \ref{fig:sfdn:surfden}, as the outflow rate
is reduced the surface density profile of the baryons more
closely resembles the initial conditions. For $t_{ml}=9$,
the baryon surface density profile is now well fit by an exponential
only interior to $\sim a_h$, with a break in the exponential profile
at this radius. Notice, however, that the analytic
calculation from appendix \ref{sec:surfdencalc} (dashed line) still
provides a good fit.

The baryon velocity anisotropy and velocity
dispersion for run C5 (see figure \ref{fig:sfdn:surfden})
follow the same trend as the baryon surface 
density: as the outflow time increases, the anisotropy introduced by
the mass loss is reduced, the drop in $\beta(r)$ moves to smaller
radii and the projected velocity dispersion becomes smoother (although
the feature corresponding to the drop and subsequent rise in
$\beta(r)$ remains).

\subsection{Why do exponential surface density profiles form?}
As demonstrated above, where mass loss significantly and rapidly
alters the central potential, approximately exponential surface
density profiles form after mass loss for a wide range of initial
conditions. The move from isotropic to radial orbits for the baryons
represents half of the answer to why this occurs. But questions remain:
what determines the final energy distribution of the baryons
after mass loss; and why, over many different initial conditions does
this lead in all cases to approximately exponential surface density
profiles? 
 
Rapid changes in the potential of a galaxy will lead to violent
relaxation in the remaining stars and dark matter, which causes the
system to move towards a Maxwellian distribution (with temperature
proportional to mass) irrespective
of the initial conditions \citep{1967MNRAS.136..101L}. It may be
simply shown that the surface density profile of a Maxwellian
distribution embedded in a Hernquist potential is approximately
exponential over a wide range of radii (see appendix
\ref{sec:violentrelax}). Furthermore, while Maxwellian distributions
are isotropic and the baryons after mass loss here are clearly not, it
may be demonstrated that the inclusion of anisotropy destroys the
exponential-like surface density profiles only if the final distribution of
baryons becomes strongly anisotropic at all radii. This is not the case
in these simulations (see also appendix \ref{sec:violentrelax}).

We conclude that the ultimate cause of the exponential-like surface
density profiles in these models is the violent relaxation of the
surviving baryons after mass loss and the move towards a
Maxwellian distribution. At large radii beyond $\sim a_h$, where the
baryons are highly anisotropic after mass loss and do not reach a
Maxwellian distribution, a second population
forms, causing a break in the surface density profile and a drop in the
projected velocity dispersion near the halo scale length.

\section{Discussion}\label{sec:discussion}

In this section we discuss the simulation results in relation to other
works on mass loss already in the literature and demonstrate that this
study complements and expands on the existing material. We also
highlight and clarify the assumptions inherent in this work.

\subsection{Comparison with previous work}
\citet*{2002MNRAS.333..299G}, \citet{1996MNRAS.283L..72N} and
\citet{1999MNRAS.303..321G} have all looked at the effect on a
Hernquist halo density profile of removing a baryon disc. The crucial
differences with our study here are as follows:\\
\\
1. The main contribution of this work is to analyse the baryons left
behind after mass loss, quantifying their angular momentum,
surface density profile and velocity structure. This then allows us to
compare the galaxy dynamical evolution models presented here with
observations (see section \ref{sec:observations}). This has not been
covered by previous authors.\\
\\
2. We start with both spherically symmetric pressure supported and
axisymmetric rotationally supported baryonic profiles as options for
the progenitors of dwarf spheroidal galaxies. The earlier studies
start with a disc with some net angular momentum. As such, the
collapse factor in the models presented 
here is a free parameter, whereas in the previous literature, it is
constrained by the initial angular momentum profile of the baryons (although
\citet{2002MNRAS.333..299G} do discuss angular momentum-free collapse).\\
\\
3. We simulate the contraction of the central dark matter
cusp and compare this with the adiabatic, analytic
prediction. \citet{2002MNRAS.333..299G} impose the analytic result on
their initial conditions for the simulations, while
\citet{1999MNRAS.303..321G} and \citet{1996MNRAS.283L..72N} use only a
numerical cusp contraction method. As demonstrated in
section \ref{sec:resdarkmatter}, while non-adiabatic inflow can often
produce density profiles which are well fit by the adiabatic
form, the non-isotropic velocity distribution created by the inflow
leads to shallower density profiles after mass loss. Thus it is
important to consider the case of faster mass inflow as compared with
the true adiabatic case.\\
\\
4. We use this numerical cusp contraction method to study the
effect of {\it multiple} mass loss phases punctuated by periods of
slow accretion. This can produce cores from cuspy dark matter
distributions, even with cosmological initial conditions.\\
\\
5. We perform fully 3D N-body simulations with non-rigid disc and halo
potentials. \citet{1999MNRAS.303..321G} do the same, but at a lower
resolution (40,000 particles). \citet{1996MNRAS.283L..72N} use a live
halo but a rigid disc potential, and again at lower resolution
(10,000 particles). Finally \citet{2002MNRAS.333..299G} use a 1D shell
code for their simulations. This has the advantage that it does not
require any force softening, but the disadvantage that it is only 1D
(and so, for example, the velocity distribution is necessarily
isotropic at all times).\\
\\
\citet{2002MNRAS.333..299G} found that mass loss from
the baryons produced a much smaller effect on the central density
profile of the dark matter than was found by
\citet{1996MNRAS.283L..72N} and \citet{1999MNRAS.303..321G}. They
suggest that this could be due to resolution, but the differing nature
of the codes employed made these comparisons difficult. In order to
shed some light on this issue, we re-simulated at {\it lower}
resolution, with 10,000 particles, the run B1(c), where the baryons
were contracted analytically and then 95\% were removed very
quickly. We found that, despite the larger Poisson errors, the results
were consistent with the higher resolution run: no core was found to
form in the dark matter central density after mass loss. It seems that
another difference 
between the studies of \citet{2002MNRAS.333..299G} and
\citet{1996MNRAS.283L..72N} must be acting: we suggest that this is
the baryon dissipational inflow rate. Both
\citet{1996MNRAS.283L..72N} and \citet{1999MNRAS.303..321G}
numerically contract their dark matter cusps. As demonstrated in
section \ref{sec:resdarkmatter}, this leads to shallower cusps after
mass outflow. The differences between these previous studies in the
literature is then likely to be a combination of numerical resolution
effects and the different methodologies which they assumed for the mass
inflow phase for the baryons.

This view is further supported by \citet{2002ApJ...571L..89J} who
looked at the effects of non-adiabatic inflow on the contraction of a
dark matter cusp. They find, similarly to the results presented here,
that fast inflow can lead to shallower contraction than the adiabatic
prediction, although they do not analyse the orbital structure of the
dark matter before and after mass inflow. This last point is critical
since density profiles which {\it appear} to be adiabatic after mass
inflow can have anisotropic velocity distributions which lead to
an increased response after mass loss.

\begin{figure*} 
\begin{center}
  \epsfig{file=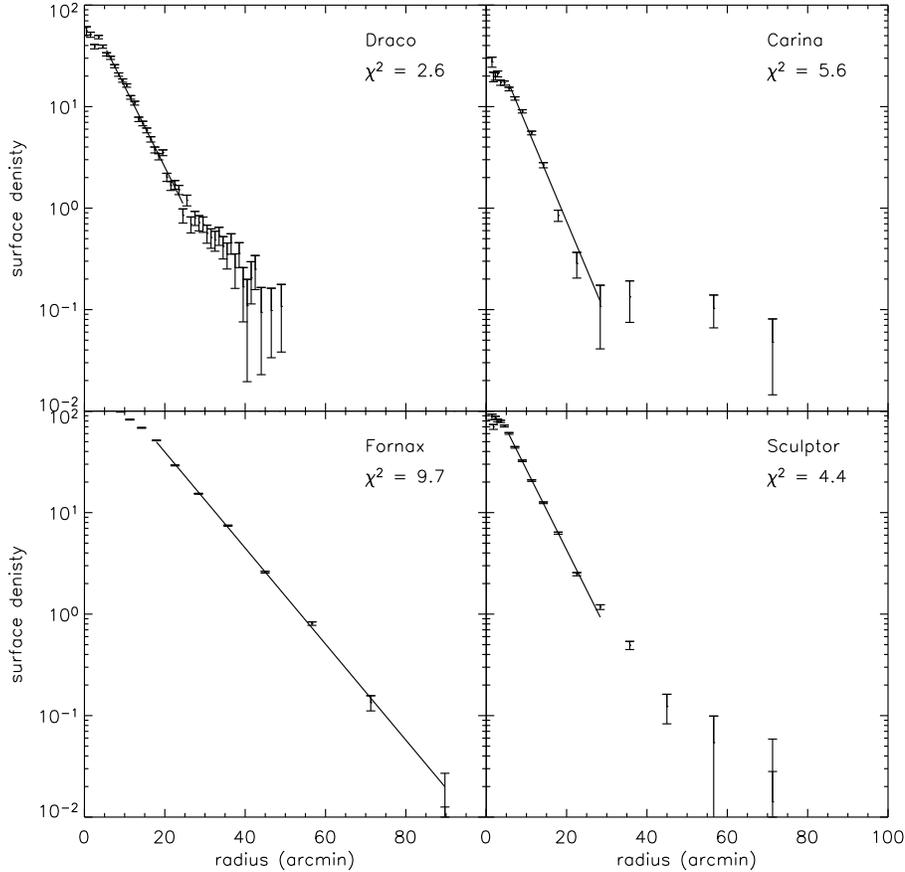,width=120mm}
  \caption[The surface density profiles of four Local Group dSph galaxies]
  {The surface density profiles of four Local Group dSph
  galaxies. The solid lines are exponential fits to the surface
  density profiles over a range of radii. $\chi^2$ for each of the
  fits is marked in the top right hand corner of each plot. The data
  were taken from Irwin, private correspondence (Draco) and
  \citet{2000Astro-Ph..0207467} (Carina, Fornax and Sculptor).}
\label{fig:localgroup}
\end{center}    
\end{figure*}

\subsection{Model assumptions}
Before discussing the results and their relation or otherwise to real
dSph galaxies, it is important to recall and to justify the
assumptions we have made both explicitly and implicitly in this
study. We list the main assumptions below and discuss them in turn:\\
\\
1. We assumed that dSph galaxies comprise a baryonic component and a
dark matter component and that the galaxy, before mass loss, is in
virial equilibrium.\\
\\
2. We assumed that the dark matter component of the galaxy is spherical
and symmetric and can be parametrised by a Hernquist
profile or a truncated isothermal sphere. Furthermore we assumed that
the initial velocity structure of the halo before baryonic mass infall
is isotropic.\\
\\
3. We assumed that the baryonic component of the galaxy can be
parametrised by a spherical, generalised Hernquist profile, or by a
rotationally supported exponential disc.\\
\\
The important assumption here is in the initial conditions used: an
isotropic galaxy in equilibrium. This is perhaps unlikely to be the
case for a real proto-galaxy. However, in terms of
the magnitude of any feedback due to mass loss from the baryons,
starting with equilibrium initial conditions represents the extremum
case. This is because this allows the baryons to collapse and
dominate the central potential {\it before} any mass loss occurs. Mass
loss which occurs at the same time as baryonic collapse would
produce weaker effects than those discussed here, all other things
being equal, since the baryons would be less centrally concentrated at
the point of mass loss.

The assumption of initial isotropy is also critical. As discussed
above, strongly radially anisotropic velocity
distributions become more easily unbound as a result of rapid changes
in the local potential (such as is caused by fast mass loss). Thus
initial radial anisotropy would {\it increase} the dynamical effects of mass
loss on the remaining stars and dark matter.

\section{Comparison with observations}\label{sec:observations}

\subsection{Angular momentum}

In section, \ref{sec:resangmom}, we considered the effect of mass loss
on the angular momentum distribution in dwarf galaxies. The best
studied Local Group dSph galaxy, Draco, is not rotating at all at the
3-sigma level \citep{2001ApJ...563L.115K}; the other Local Group
dSph galaxies also show near complete pressure support, with very
little rotation (see e.g. \citet{1998ARA&A..36..435M}). However, if
the progenitor galaxies to dSph at the present epoch are consistent
with scale-free $\Lambda$CDM cosmology, then they should form from gas
with significant global internal angular momentum (see
e.g. \citet{2001ApJ...555..240B} and \citet{Colin:2003jd}). Could
random or biased mass loss explain this angular momentum discrepancy?

Only three simulations produced significant angular
momentum loss. The first, run A6, had almost no dark matter initially
interior to the baryons, while the initial baryon distribution was
mostly in pressure support, with $v_r/\sigma = 0.8$. This run, which
represented the most extreme case for random mass removal, was then
already a low-angular momentum system before mass loss. 

The other simulations which produced significant angular momentum loss were
the biased mass loss runs: runs A7 and A8. These preferentially
removed mass at large radii, thus removing much more of the initial angular
momentum for a given mass loss fraction. While this type of mass loss
would not be caused by supernovae explosions which would be expected to
remove preferentially the {\it low} angular momentum gas (see
e.g. \citet{2002MNRAS.335..487M}); this type of mass loss could
correspond to strong tidal stripping caused as a proto-dSph galaxy
passes close to a larger nearby galaxy (see
e.g. \citet{2001ApJ...547L.123M}). However, both runs A7 and A8 left
remnants which had a larger final angular momentum than that measured
in the Draco dSph galaxy \citep{2001ApJ...563L.115K}.

It is clear that in order to model the angular momentum
distribution of the Local Group dwarf galaxies, these galaxies must be
simulated in a cosmological context where the timescales for gas
dissipation, dark matter collapse, merging and feedback are all
comparable. However, the results from this study suggest that the
origins of low angular momentum {\it tidally isolated} systems are not
a result of gas mass loss. Furthermore, even biased mass loss
scenarios such as tidal stripping would leave a measurable amount of
angular momentum behind if dSph galaxies really formed from
high-angular momentum gas discs.

\subsection{The effect of mass loss on the central dark matter
  distribution}

In section, \ref{sec:resdarkmatter}, we considered the effect of mass
loss from the baryons on the underlying dark matter distribution. In
particular we considered whether or not a dark matter cusp with log
slope, $\alpha={-1}$, such as is predicted by cosmological simulations of
dark matter halo formation, could survive one or many phases of mass
loss from the baryons. On the smallest scales, from the Milky Way to
dIrr and LSB galaxies, 
the central dark matter density profile is well fit inside $\sim 1$kpc
by an approximately 
constant density {\it core} (see e.g. \citet{2001ApJ...552L..23D},
\citet{2002Astro-Ph..0203457} and \citet{2001MNRAS.327L..27B}), rather than
the $1 < \alpha < 1.5$ log-slope predicted by current cosmological
simulations (see e.g \citet{2001ApJ...554..903K},
\citet{2000ApJ...544..616G} and
\citet{1996ApJ...462..563N}). Furthermore, there is recent 
evidence that this problem persists on the smallest scales, within
dSph galaxies \citep{2003ApJ...588L..21K}. It is interesting, then, to
consider whether or not extreme mass loss from the baryons can resolve
the cusp-core problem by lowering the central dark matter density.

In all of the simulations in section \ref{sec:resdarkmatter} the dark matter
responded dynamically to the time dependent potential, provided that
baryons dominated the central 
potential initially, and provided that the mass loss was rapid (of the
order a baryon crossing time). However, if the dark matter was allowed
to contract adiabatically in response to the slow dissipation and
collapse of the 
baryons, the final effect after mass loss on the density profile of
the halo was small. Reducing the inflow time of the baryons and moving
away from the adiabatic regime produced shallower final density
profiles after mass loss but, for cosmologically consistent initial
conditions, the 
final halo density profiles retained central slopes which were close
to the initial conditions - although with lower central normalisations
(see for example runs B1 and B4). Only for the two extreme initial
condition simulations (B2 and B3), did the final halo density profile exhibit
a shallower central density {\it slope}. The similarity in this
respect between these two simulations is not surprising: they both
had a very low central mass of dark matter before mass loss as
compared with the mass of the baryons. For run B2, this was achieved
by increasing the halo scale length and moving all of the mass out to
larger radii. For run B3, this was achieved by reducing the total halo
mass to be just two times the total mass of the baryons initially. 

While these extreme simulations can transform central dark matter density
cusps into cores, neither of them provides a neat solution to the
cusp-core problem. This is because for run B2, the scale length of the
halo is so large initially, ($a_h/h_b=50$), that
it must either correspond to a galaxy much smaller than present-day
dSph galaxies, or correspond to the formation of very extended
low-concentration halos, which are not predicted to form in
cosmological simulations. Put more quantitatively, cold dark matter
 simulations predict that halos forming on the smallest
scales should have concentration parameters greater than $\sim 10$
\citep{1996ApJ...462..563N}, which fixes the collapse factor to be
{\it at  most} $a_h/h_b \sim 10$ for a halo of mass
$10^{10}$M$_\odot$. Thus, although there is no current {\it observational}
constraint on the collapse parameter, there is a self-consistency
constraint from numerical simulations. Similarly, run B3 requires an
initial baryon fraction of 
$f_b = 0.33$ which is much larger than the cosmological mean value of
$f_b=0.17$ \citep{2003Astro-Ph..0302209}. This implies a
situation where a halo accumulates an initial over-abundance of baryons,
then loses a very significant amount of its baryonic
mass. The baryonic mass of a progenitor to such a dSph galaxy
would be some $\sim 10^8$M$_\odot$, assuming a current baryonic mass
of $\sim 10^6$M$_\odot$ \citep{2001ApJ...563L.115K}.

While single mass loss events do not significantly perturb the
underlying dark matter distribution, {\it multiple} mass loss 
events can cause significant flattening of the central dark matter
density for initial conditions which represent the cosmological mean
(see run B4). This is due to the iterative asymmetry 
introduced by a slow, {\it but non-adiabatic} inflow followed by a
fast, impulsive, mass loss. Thus, if dSph galaxies have undergone two
or more violent star formation bursts, with rapid baryon re-accretion
then their central dark matter density profile could have been
transformed from a $r^{-1}$ cusp into a $\sim r^{-0.2}$ core.

\subsection{The baryon surface density profile}

Figure \ref{fig:localgroup} shows a plot of the surface
density profiles of four Local Group dSph galaxies (data were taken
from \citet{Wilkinson:2004fz} and 
\citet{2000Astro-Ph..0207467}). Notice that all of the profiles, over a
wide range of radii, are reasonably well fit by an exponential -
especially the Draco dSph galaxy ($\chi^2$ for the fits is shown in
the top right hand corner of each plot). Notice further that three of
the four profiles show a clear break in the surface brightness profile of
the light at what may be a tidal radius (see
e.g. \citet{2000Astro-Ph..0207467}, \citet{1998ARA&A..36..435M} and
\citet{2003AJ....125.1926G}). The one galaxy which does not
show such a break, Fornax, is also the galaxy furthest from the Milky
Way at present \citep{1998ARA&A..36..435M}.

These profiles are qualitatively well fit by the simulation
profiles in which exponential-like profiles form over a wide range of radii
in all cases. In fact the `tidal feature' could also be the result of
mass loss. A break in the surface density profile was observed in all
of the simulations either as a result of the long-time evolution
of the baryons (see run C1), or as a result of slower (run C5) or less
strong (runs C3 and C4) mass loss.

This is particularly interesting since several authors have proposed
that the Local Group dSph galaxies may reside in much larger dark matter
halos (see e.g. \citet{2003ApJ...584..541H}). If this were the case then
their tidal radii would lie far beyond the luminous matter and the
observed `tidal features' in the surface density profiles would 
require another explanation. We propose here that one explanation for
these features could be mass loss.

Finally, these results may scale up to larger
elliptical galaxies. \citet{2004ApJ...600L..39T} have recently shown
that there is a strong correlation between the surface brightness profiles of
elliptical galaxies and their absolute magnitudes: lower magnitude
elliptical galaxies have more exponential-like surface density
profiles. If the 
absolute magnitude of ellipticals is a good tracer of their total mass
then this correlation could be explained by gas mass
loss driven by feedback - either from star formation or from a central
active nucleus. Feedback would then be more effective in less massive
systems and so less massive, lower luminosity elliptical galaxies
would have more exponential-like surface density profiles, as
observed.

\subsection{The baryon projected velocity dispersion}

While the detailed form of the projected velocity dispersion after
mass loss varied for different initial conditions, there was one
common feature: the formation of a new, radially-biased,
population of baryons beyond the original distribution after mass loss
led, in all cases, to a drop in the projected velocity dispersion at
$\sim a_h$, the halo scale length.

This is interesting since recent results from \citet{Wilkinson:2004fz}
find just such a drop in the projected velocity dispersion for the
Draco and UMi dSph galaxies.

The good qualitative agreement between both the surface density
profiles and projected velocity dispersions (where measured) of the
Local Group dSph galaxies and the dynamical models presented in this
paper indicates that mass loss may be the primary evolutionary factor
in determining the structure of dSph galaxies at the present epoch. To
test this idea further, a next step towards building a more realistic
model for the formation of the Local Group dSph galaxies should
include the effect of an external tidal field from the proto-Milky
Way; this will provide the subject of future work.

\section{Conclusions}\label{sec:conclusions}

We have looked at the dynamical effect of baryonic mass loss from
two-component galaxies comprising a dark matter halo and some baryons,
initially in equilibrium. These initial conditions represent an
extreme case; by studying this we have been able to quantify the
potential importance of mass loss on astrophysically interesting observational
signatures, such as the central dark matter density, 
internal angular momentum distribution and baryon surface density profile
of these galaxies. 

Our focus was on simulating the effects of mass loss on progenitors to
what are currently dwarf spheroidal (dSph) galaxies. This is because
these galaxies appear observationally to have lost a large fraction of
their gas mass at some point in the past.

Our main findings are summarised below:\\
\\
1. One impulsive mass loss event involving a sufficiently
large fraction of the baryons will naturally produce a surface density
profile in the remaining stellar component of a dSph which is well fit
by an exponential over many scale lengths, in good qualitative agreement with
observations of the Local Group dSphs. Furthermore, in nearly
all models, a break is naturally found to form in the surface density
profile at large radii, which could be easily
mistaken for a feature of tidal origin. This lends support to the idea
that the true tidal radius of the Local Group dSph galaxies may lie
well beyond the observed light.\\
\\
2. Mass loss naturally leads to a drop in the projected velocity
dispersion of the light at around the halo scale length, also in good
qualitative agreement with recent observations of the Draco and UMi
dSph galaxies.\\
\\
3. Two impulsive mass loss phases, punctuated by significant gas
re-accretion, are found to be sufficient to transform a central
density cusp in the dark matter profile into a near-constant density
core. This may then provide the missing link between current
cosmological simulations, which predict a central cusp in the dark
matter density profile, and current observations, which find much
shallower central density profiles.\\
\\
4. Mass loss cannot account for the currently observed low angular
momentum of the Local Group dSph galaxies. Thus, if local group dSph
galaxies have spent most of their lifetime in tidal isolation from
massive galaxies, then they cannot have formed from high angular
momentum gas discs.

\section{Acknowledgements}
We would like to thank Lars Hernquist for his galaxy initialisation
code and Jakob Walcher for the data for the Carina, Fornax and Sculptor
galaxies. Finally we would like to thank Mark Wilkinson and the
referee for useful comments which led to the completion of this
manuscript. Part of this work was conducted on the SGI Origin platform
using COSMOS Consortium facilities, funded by HEFCE, PPARC and SGI.

\vfil
\bibliographystyle{apj_jss}
\bibliography{refs}

\appendix
\section{The response of a dark matter halo to gas collapse and gas mass
loss}\label{sec:zhaocalc}

In this appendix we briefly outline an analytic calculation to
determine the response of a Hernquist dark matter halo (see equation
\ref{eqn:hernquist}) to the adiabatic dissipation and collapse and
subsequent mass loss of some baryons. The calculation is similar to
that set out in \citet{2002MNRAS.333..299G}, but using a pressure
supported Hernquist profile for the initial conditions for the
baryons. As such, we highlight only the differences here.

\subsection{The baryon dissipation and collapse phase}\label{sec:analcont}

To model the adiabatic contraction of the central dark matter, we
assume that the baryons and the dark matter start out with the same
distribution - a Hernquist profile (see equation
\ref{eqn:hernquist}) and that the dark matter particles lie on a
collection of randomly oriented circular orbits. The baryons then
collapse to form a smaller Hernquist profile, with scale length $h_b$
and mass $M_b$.

If the collapse conserves both the angular momentum of a dark matter
particle's orbit and the dark matter mass interior to its orbit then it
can be shown that:

\begin{equation}
\frac{C}{b}a^3 + (\frac{2C}{b}-1)a^2 + (\frac{C}{b}-2)a - (1+K(1+b)^2)=0
\label{eqn:efdm:cubic}
\end{equation}
\noindent
where 

\begin{eqnarray}
C & = & \frac{a_h}{(1-f_b)h_b} \label{eqn:efdm:C} \\
K & = & \frac{f_b}{1-f_b} \label{eqn:efdm:K}
\end{eqnarray}

\noindent
and $r_i$ and $r_j$ are the pre and post-collapse radii; $a_h$ is the
initial baryon scale length before collapse, which is also the halo
scale length; $M_h$ is the halo mass; $f_b = \frac{M_b}{M_b+M_h}$ is
the baryon fraction; $a=h_b/r_j$ and $b=a_h/r_i$. 

The final mass distribution, $M_{dm,j}(r_j)$, and density distribution
can then be found by solving equation \ref{eqn:efdm:cubic} numerically
and using mass conservation.

\subsubsection{Adiabatic and instantaneous winds}\label{sec:efdm:adiimpwinds}
We follow the prescription in
\citet{2002MNRAS.333..299G} and use an heuristic formula for the
impulsive mass loss case. The formula results from considering the
limiting cases of adiabatic mass loss and the
limiting case of impulsive mass loss, where it is assumed that the
system becomes unbound if it instantly loses half of its baryonic mass
\citep{2002MNRAS.333..299G}. As shown in section \ref{sec:resdarkmatter},
this formula provides a good fit to the simulation data in the
appropriate limits. Following \citet{2002MNRAS.333..299G}, then, we
have: 

\begin{equation}
r_f =  \frac{r_j}{(1-f_W(r_j))^{1+\frac{k}{4}}(1-2^k\delta
 f_b)^{\frac{k}{4}}}\label{eqn:efdm:genwind}
\end{equation}
\noindent
where
\begin{equation}
f_W(r_j) = \frac{\delta M_b(r_j)}{M_{dm}(r_j)+M_b(r_j)}
\end{equation}
\noindent
and $0 \leq k \leq 1$ is a tunable, free parameter, to model the
rapidity of the wind. 

It is important to note here that this heuristic argument can be
expected to break down for extreme initial conditions. For example,
\citet{2003MNRAS.338..665B} and \citet{2003MNRAS.338..673B} have
recently shown that the fraction of stars which remain bound after an
impulsive mass loss is generally higher than standard virial arguments
would suggest.

\subsubsection{Non-adiabatic inflow models}\label{sec:efdm:minheating}
So far, we have considered only the special case of adiabatic gas
infall. We also present some
simulations in which this assumption is relaxed. As such, it is
instructive to attempt to consider analytically how we might expect
the cusp contraction to change when gas is dropped on some shorter
timescale, $t_{cont}$. To do this, it is helpful to consider the
energy of a dark matter particle's orbit before and after the mass
infall. If we start with the same initial conditions before mass
infall as in section \ref{sec:analcont}, then all of the particles
start on circular orbits. Thus the initial specific energy,
$E_{dm,i}$, of a dark matter particle at a radius, $r$, is given by: 

\begin{equation}
E_{dm,i} = \frac{1}{2}v_{dm,i}(r)^2 + \Phi_{dm,i}(r)
\label{eqn:efdm:ei}
\end{equation}

where $v_{dm,i}(r)$ is the circular velocity of the particle at $r$ and
$\Phi_{dm,i}(r)$ is the local potential. After the mass infall, the local
potential is increased by the addition of the baryons. For adiabatic
changes, where circular orbits are maintained, this gives for the
specific energy after mass infall, $E_{dm,f}^*$:

\begin{equation}
E_{dm,f}^* = \frac{1}{2}v_{dm,f}(r)^2 + \Phi_{dm,f}(r) + \Phi_b(r)
\label{eqn:efdm:efcirc}
\end{equation}

where the $*$ denotes circular orbits, and $\Phi_b(r)$ is the
baryonic contribution to the local potential.

Since angular momentum will be conserved throughout the mass inflow,
circular orbits are then the lowest final energy state of the
system. For more general initial conditions, one can think of the dark
matter particles being {\it heated}. The adiabatic case, which
preserves the initial orbital structure, is then the situation of {\it
  minimal heating}. Thus the adiabatic case produces the {\it lowest}
final specific orbital energy for the dark matter after mass infall.

For non-adiabatic mass infall, the final specific energy of a dark
matter particle, $E_{dm,f}$, will then be higher than for the adiabatic
case:

\begin{equation}
E_{dm,f} = E_{dm,f}^*\delta_E
\label{eqn:efdm:ef}
\end{equation}

where $0 < \delta_E \le 1$ essentially parameterises our ignorance
about the final energy state of a dark matter particle. Notice that
for $\delta_E = 1$ we recover circular orbits and adiabatic mass
infall. For $\delta_E < 1$, the final energy state is {\it higher} in
the sense that it is a smaller negative number, than for the adiabatic
case. For non-adiabatic changes, some of the energy that would have
gone into contracting the central dark matter cusp now 
goes into heating the central orbital structure. This produces {\it
  two} important effects: firstly, an initially isotropic velocity
distribution for the dark matter will be heated and become more
radially anisotropic in the centre. This is important since particles
may then become more easily unbound after a subsequent mass
loss. Secondly, the contraction of the halo can be expected to be less
for non-adiabatic inflow than for the adiabatic case.

\section{The response of a stellar distribution to impulsive gas mass loss}\label{sec:surfdencalc}

In this appendix we briefly summarise an analytic calculation to
determine the response of the remaining baryons to an impulsive mass
loss from the baryons. The approach is similar to the orbit
integration method proposed by \citet{1979ApJ...232..236S} for setting
up equilibrium galaxies; however, the context is very different.

\subsection{Impulsive mass loss}\label{sec:sfdn:analimp}
We use initial conditions before mass loss as in section
\ref{sec:initialcond}, but we place all of the baryon particles on
randomly oriented circular orbits. For here on, `particle' refers to a
baryon or star particle.

In the impulsive regime, after the mass loss, particles which were
initially on circular orbits with specific energy, $E_i$, at a
radius, $r_i$, then move onto some more general orbit with specific
energy, $E_f$, at a radius, $r_f(t)$, which is now a {\it function of 
  time}. For general orbits in a spherically symmetric potential, it
can be shown that $r_f(t)$ obeys the equation
\citep{1987gady.book.....B}:

\begin{equation}
\left(\frac{dr_f}{dt}\right)^2 = 2(E_f - \Phi_f) - \frac{L^2}{r_f^2}
\label{eqn:sfdn:orbiteq}
\end{equation} 

where $\Phi_f(r_f)$ is the potential of the baryons and dark matter after
the mass loss and $L(r_i)$ is the angular momentum per unit mass of the
orbit, which is fixed by angular momentum conservation to be the same
as the angular momentum of the initial circular orbit at $r_i$.

Equation \ref{eqn:sfdn:orbiteq} has, for bound orbits, in general, two
solutions where $\frac{dr_f}{dt}=0$.

Thus, if $\Phi_f(r_f)$, $E_f(r_i)$ and $L^2(r_i)$ are known, we can calculate
$r_{max}$ and $r_{min}$ from equation \ref{eqn:sfdn:orbiteq}. We can
then calculate the final density profile of the baryons, $\rho_{b,f}$,
in the following way: rearranging equation \ref{eqn:sfdn:orbiteq}, we
can integrate to find the total time, $t_{orb}$, a particle spends moving from
$r_{min}$ to $r_{max}$. This gives: 

\begin{equation}
t_{orb}(r_i) = \int_{r_{min}}^{r_{max}}{\frac{dr_f}{\sqrt{2(E_f(r_i)-\Phi_f(r_f)) - \frac{L(r_i)^2}{r_f^2}}}}
\label{eqn:sfdn:torb}
\end{equation} 

Furthermore, we can similarly calculate the time spent by the particle
in some small, infinitesimal, radial slice between $r$ and
$r+dr$. This gives:

\begin{eqnarray}
t_{slice}(r,r_i) & = &
      \int_{r}^{r+dr}{\frac{dr_f}{\sqrt{2(E_f(r_i)-\Phi_f(r_f)) -
      \frac{L(r_i)^2}{r_f^2}}}} \nonumber \\
& = & dr\left(2(E_f(r_i)-\Phi_f(r_f)) -
      \frac{L(r_i)^2}{r^2}\right)^{-1}
\label{eqn:sfdn:tslice}
\end{eqnarray} 

The {\it probability} of a particle initially at a radius,
$r_i$, being at a radius, $r$, is then given by:

\begin{equation}
p(r,r_i) = \frac{t_{slice}(r,r_i)}{t_{orb}(r_i)}
\label{eqn:sfdn:prri}
\end{equation} 

So, to recapitulate: a particle initially at a radius, $r_i$, will
move onto a time 
dependent orbit, $r_f(t)$, which samples a {\it range} of radii between
$r_{min}$ and $r_{max}$. The final mass distribution at some radius
$r$ of many particles will then be the statistical sum of the probabilities of
finding each of the particles at $r$. Thus, if the initial density
distribution of baryons is given by $\rho_{b,i}$, then the final
density distribution will be given by:

\begin{equation}
4\pi r^2\rho_{b,f}(r)dr = \int_{0}^{\infty}{p(r,r_i)4\pi r_i^2 \rho_{b,i}(r_i)dr_i}
\label{eqn:sfdn:rhor}
\end{equation} 

The term on the left hand side is in fact the final {\it mass} of an
infinitesimal shell of radius, $r$ and thickness, $dr$. However, this
can be numerically differentiated to find the density profile and then
numerically integrated to find the surface density profile.

The final potential after mass loss, $\Phi_f(r_f)$, may be calculated
as in appendix \ref{sec:zhaocalc} if it is assumed that the baryons
contribute negligibly to the final potential. Thus, the only unknown
in equation \ref{eqn:sfdn:tslice} is the final specific energy distribution of
the baryon orbits, $E_f(r_i)$. In the adiabatic, or infinitely slow
mass loss case, where the orbital structure is preserved, this may be
simply calculated. In general, however, this value must be taken from the
simulations.

\section{Violent relaxation in a Hernquist potential and the formation
  of exponential surface density profiles}\label{sec:violentrelax}
\citet{1967MNRAS.136..101L} demonstrated that violent relaxation will
cause particles to move towards a Maxwellian distribution (with
temperature proportional to mass) irrespective of the initial
conditions. Where relaxation ceases before equilibrium is reached; or
where a non-Maxwellian steady state solution is reached first, the
system should still be close to Maxwellian
\citep{1967MNRAS.136..101L}. Furthermore the system is more likely to
achieve isotropy at small radii, where dynamical times are short, than
at larger radii. As such, we follow
\citet{1979PAZh....5...77O} and \citet{1985MNRAS.214P..25M} and
parameterise the final post-mass loss non-isotropic baryon
distribution function by the following form:

\begin{equation}
f = Ae^{-\beta(E+L^2/{2r_a^2})}
\end{equation}
\noindent
where $E$ is the specific energy of the baryons, $L$ is the specific
angular momentum and $r_a$ is the radius at which the velocity of the
baryons moves from being isotropic to being anisotropic
(\citet{1979PAZh....5...77O} and \citet{1985MNRAS.214P..25M}). Notice
that for $r_a \rightarrow \infty$ we recover the true Maxwellian
distribution at all radii.

The density of the baryons after mass loss may then be calculated from
$f$ as in \citet{1987gady.book.....B} to give: 

\begin{equation}
\rho(\psi,r) = \frac{2\sqrt{2}\pi A
(\sqrt{\pi}e^{\beta \psi}\mathrm{erf}(\sqrt{\beta \psi})-2\sqrt{\beta
  \psi})}
{\beta^{3/2}(1+(r/r_a)^2)}
\end{equation}

and final surface density numerically integrated from the
above equation using the effective potential, $\psi$, for a generalised
Hernquist profile (see equation \ref{eqn:spheroid} and
\citet{1987gady.book.....B} for a definition of the effective
potential).

\begin{figure} 
\begin{center}
  \epsfig{file=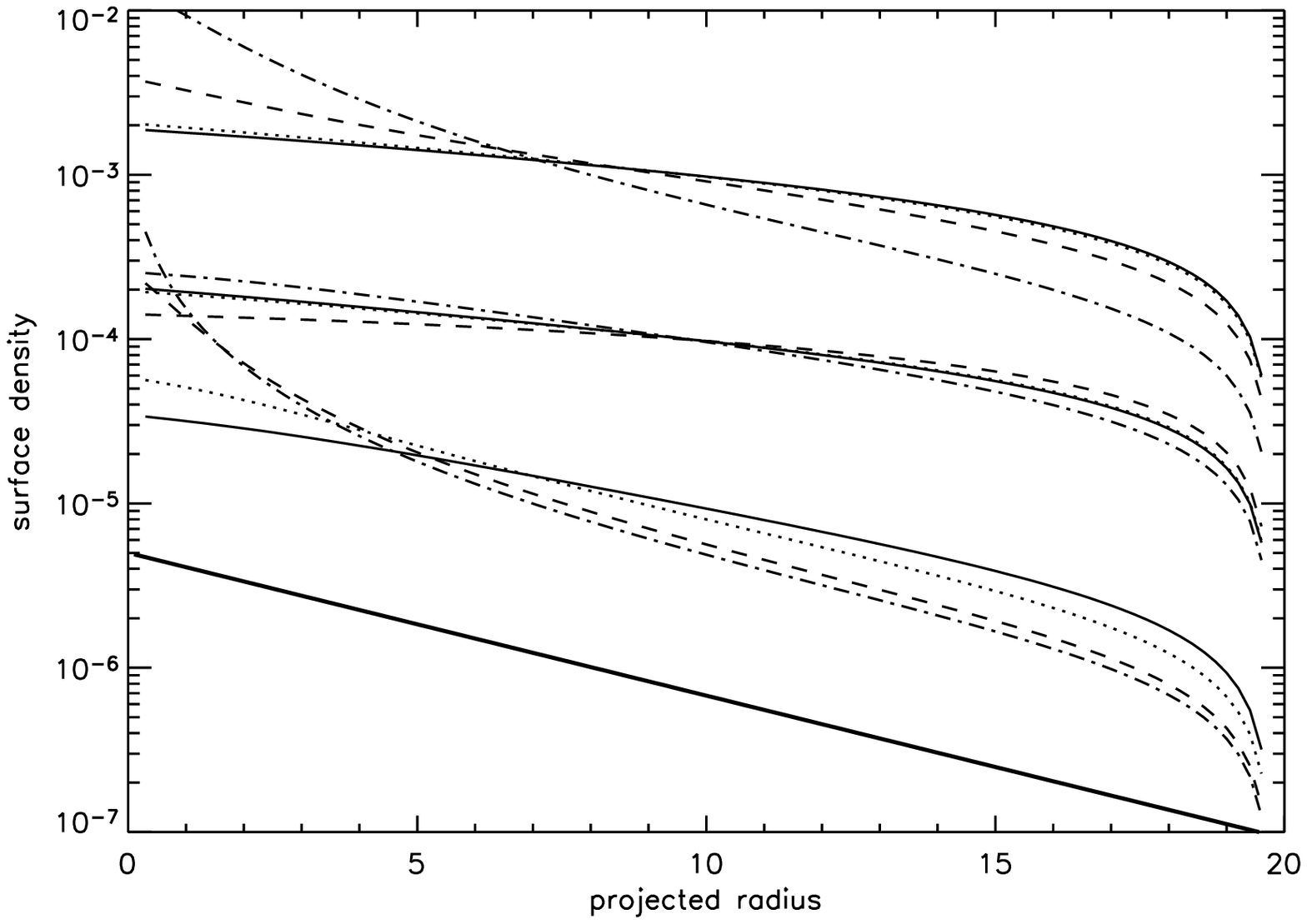,width=0.5\textwidth}
  \caption[]
  {The surface density profile of a violently relaxed stellar
  population embedded in a Hernquist dark matter potential. The curves
  have been offset from one another for clarity. From top to bottom
  the three sets of curves show: 1. the effect of varying $\beta$ for a
  fixed halo potential (the solid, dotted, dashed and dot-dashed lines
  are for $\beta=0.1,1,5$ and $10$ respectively); 2. the effect of
  varying the halo potential for fixed $\beta$ (the lines are for
  halos with the same initial conditions as models B1-B4 (see table
  \ref{tab:initcond}) and 3. the effect of varying the velocity
  anisotropy parameter, $r_a$ (the lines are for $r_a=10,5,1$ and $0.1$
  respectively). A true exponential surface density is also
  marked for comparison (thick solid line at the bottom).}
\label{fig:violent_relax}
\end{center}    
\end{figure}

Figure \ref{fig:violent_relax} shows the surface density profiles
calculated as above for a wide range of halo potential parameters, and
values of $\beta$ and $r_a$ (while $\beta$ should strictly be
calculated from the total energy of the system, we treat this here as
a free parameter). 

Notice that only two of the curves deviate strongly from an
exponential near the centre: the dashed and dot-dashed lines in the
bottom set of curves. These two curves were for $r_a=1$ and $0.1$
respectively. In both cases this represents a much more significant
central velocity anisotropy than was observed in any of the
simulations presented in this paper. As mentioned above, it should be
of no surprise than the baryons achieve approximate isotropy at small
radii where the dynamical times are short: the system relaxes fully
at these radii into a Maxwellian distribution.
\label{lastpage}

\end{document}